\documentclass{aa}  

\usepackage{graphicx}
\usepackage{txfonts}
\begin{document}

   \title{Characterising the Apertif primary beam response}


   \author{H.~D\'enes \inst{\ref{astron}}\thanks{helgadenes@gmail.com},
          K.~M.~Hess       \inst{\ref{iaa},\ref{astron},\ref{kapteyn}}
          E.~A.~K.~Adams       \inst{\ref{astron},\ref{kapteyn}},
          A.~Kutkin            \inst{\ref{astron},\ref{lebedev}},
          R.~Morganti   \inst{\ref{astron},\ref{kapteyn}},
          J.~M.~van~der~Hulst  \inst{\ref{kapteyn}},
          T.~A.~Oosterloo      \inst{\ref{astron},\ref{kapteyn}},
          V.~A.~Moss    \inst{\ref{csiro},\ref{sydney},\ref{astron}},
          B.~Adebahr           \inst{\ref{airub}},
          W.~J.~G.~de~Blok  \inst{\ref{astron},\ref{uct},\ref{kapteyn}},
          M.~V.~Ivashina   \inst{\ref{chalmers}}, 
          A.~H.~W.~M.~Coolen   \inst{\ref{astron}},
          S.~Damstra            \inst{\ref{astron}},
          B.~Hut           \inst{\ref{astron}},
          G.~M.~Loose      \inst{\ref{astron}},
          D.~M.~Lucero         \inst{\ref{virginiatech}},
          Y.~Maan          \inst{\ref{ncra},\ref{astron}},
          \'A.~Mika             \inst{\ref{astron}},
          M.~J.~Norden          \inst{\ref{astron}},
          L.~C.~Oostrum        \inst{\ref{astron},\ref{uva},\ref{escience}},
          D.~J.~Pisano        \inst{\ref{WVU},\ref{GWAC},\ref{uct}},
          R.~Smits             \inst{\ref{astron}},
          W.~ A.~van~Cappellen	\inst{\ref{astron}},
        R.~van~den~Brink \inst{\ref{TRICAS},\ref{astron}},
        D.~van~der~Schuur	\inst{\ref{astron}},
        G.~N.~J.~van~Diepen	\inst{\ref{astron}},
        J.~van~Leeuwen    \inst{\ref{astron},\ref{uva}},
        D.~Vohl             \inst{\ref{uva},\ref{astron}},
        S.~J.~Wijnholds      \inst{\ref{astron}},
        J.~Ziemke            \inst{\ref{astron},\ref{oslocit}}
          }

 \institute{ASTRON, the Netherlands Institute for Radio Astronomy, 7991 PD Dwingeloo, The Netherlands\label{astron}
 \and
Kapteyn Astronomical Institute, University of Groningen, P.O. Box 800, 9700 AV Groningen, The Netherlands\label{kapteyn}
 \and
Instituto de Astrof\'isica de Andaluc\'ia, CSIC, Glorieta de la Astronom\'ia, E-18080, Granada, Spain\label{iaa}
  \and
Astro Space Center of Lebedev Physical Institute, Profsoyuznaya Str. 84/32, 117997 Moscow, Russia\label{lebedev}
  \and
CSIRO Astronomy and Space Science, Australia Telescope National Facility, PO Box 76, Epping NSW 1710, Australia\label{csiro}
  \and
Sydney Institute for Astronomy, School of Physics, University of Sydney, Sydney, New South Wales 2006, Australia\label{sydney}
  \and
Ruhr University Bochum, Faculty of Physics and Astronomy, Astronomical Institute (AIRUB), Universit\"atsstrasse 150, 44780 Bochum, Germany\label{airub}
  \and
Dept.\ of Astronomy, Univ.\ of Cape Town, Private Bag X3, Rondebosch 7701, South Africa\label{uct}
  \and
Dept.\ of Electrical Engineering, Chalmers University of Technology, Gothenburg, Sweden\label{chalmers}
  \and
Department of Physics, Virginia Polytechnic Institute and State University, 50 West Campus Drive, Blacksburg, VA 24061, USA\label{virginiatech}
 \and
National Centre for Radio Astrophysics, Tata Institute of Fundamental Research, Pune 411007, Maharashtra, India \label{ncra}
  \and
Anton Pannekoek Institute, University of Amsterdam, Postbus 94249, 1090 GE Amsterdam, The Netherlands\label{uva}
 \and
Netherlands eScience Center, Science Park 140, 1098 XG, Amsterdam, The Netherlands
 \label{escience}
  \and
West Virginia University, White Hall, Box 6315, Morgantown, WV
26506\label{WVU}
  \and
Center for Gravitational Waves and Cosmology, West Virginia University, Chestnut Ridge Research Building, Morgantown, WV 26505\label{GWAC}
  \and
Tricas Industrial Design \& Engineering, Zwolle, The Netherlands\label{TRICAS}
  \and
University of Oslo Center for Information Technology, P.O. Box 1059, 0316 Oslo, Norway \label{oslocit}
             }

   \date{2021}
   
   \titlerunning{Apertif primary beam response}
   \authorrunning{H.~D\'enes et al.}

 
  \abstract
   {Phased Array Feeds (PAFs) are multi element receivers in the focal plane of a telescope that make it possible to form simultaneously multiple beams on the sky by combining the complex gains of the individual antenna elements. Recently the Westerbork Synthesis Radio Telescope (WSRT) was upgraded with PAF receivers and carried out several observing programs including two imaging surveys and a time domain survey. The Apertif imaging surveys use a configuration, where 40 partially overlapping compound beams (CBs) are simultaneously formed on the sky and arranged in an approximately rectangular shape.}
   {This manuscript aims to characterise the response of the 40 Apertif CBs to create frequency-resolved, I, XX and YY polarization empirical beam shapes. The measured CB maps can be used for image deconvolution, primary beam correction and mosaicing of Apertif imaging data.}
   {We use drift scan measurements to measure the response of each of the 40 CBs of Apertif. We derive beam maps for all individual beams in I, XX and YY polarisation in 10 or 18 frequency bins over the same bandwidth as the Apertif imaging surveys. We sample the main lobe of the beams and the side lobes up to a radius of 0.6 degrees from the beam centres. In addition, we derive beam maps for each individual WSRT dish as well.}
   {We present the frequency and time dependence of the beam shapes and sizes. We compare the compound beam shapes derived with the drift scan method to beam shapes derived with an independent method using a Gaussian Process Regression comparison between the Apertif continuum images and the NRAO VLA Sky Survey (NVSS) catalogue. We find a good agreement between the beam shapes derived with the two independent methods.}
   {}


   \maketitle
%

\section{Introduction}

Phased-array feeds (PAFs) are an important advancement in radio astronomy. PAFs are multi element receivers in the focal plane of a telescope that make it possible to form simultaneously multiple beams on the sky by combining the complex gains of the individual antenna elements. This significantly increases the instantaneous field of view and the survey speed of the telescope.

\begin{figure}
   \centering
   \includegraphics[width=8.4cm]{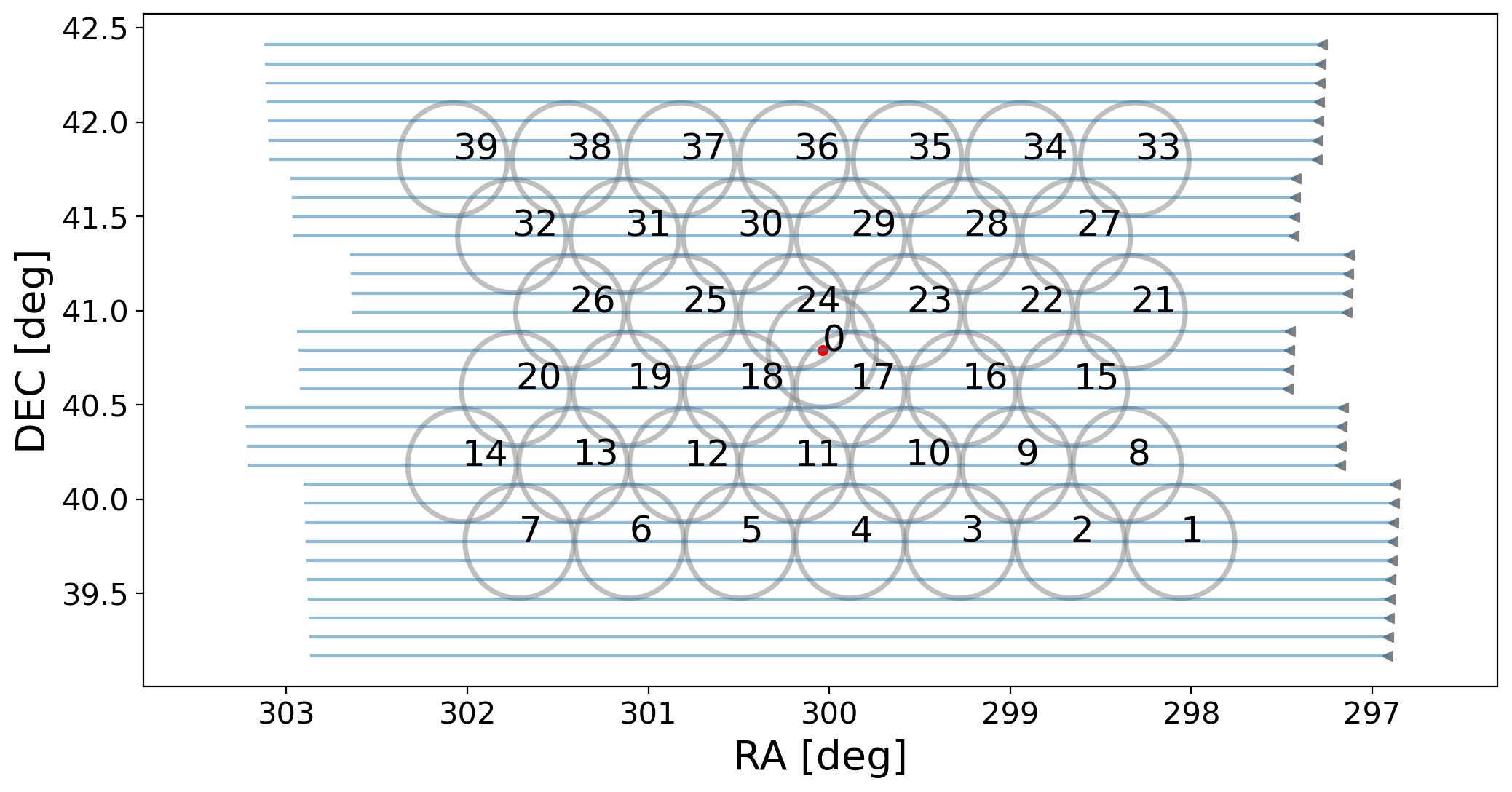}
   \caption{Illustration of drift scan observations. The red dot represents the centre of the PAF at the position of CygA. The numbered circles represent the positions of the 40 CBs (the diameter of the circles is 0.6 deg). The blue lines represent the 33 individual drifts across the field of view of the Apertif footprint, where the grey triangles mark the starting position of each individual drift.}
              \label{fig:driftscan}%
    \end{figure}

In the last few years several telescopes were equipped with PAF receivers, from single dishes like the Robert C. Byrd Green Bank Telescope (GBT; e.g. \citealt{Landon2010, Roshi2018, Pingel2021}), the Effelsberg telescope (e.g. \citealt{Chippendale2016, Reynolds2017, Deng2017}), to interferometers like the Australian SKA Pathfinder Telescope (ASKAP; e.g. \citealt{Hotan2014, Hotan2021, McConnell2016}), and the Westerbork Synthesis Radio Telescope (WSRT), which was equipped with a PAF system called Apertif (APERture Tile In Focus). Out of the 14 WSRT dishes 12 were upgraded with the new PAF receivers. Each of the PAFs has 121 Vivaldi antenna elements which can be combined to form multiple beams simultaneously on the sky. In the standard observing mode for imaging 40 partially overlapping compound beams (CBs) are formed in a rectangular configuration (Figure~\ref{fig:driftscan}). The Apertif system is tunable to observe between 1130-1750 MHz in a 300 MHz wide band. The observing band is subdivided into 384 781.25 kHz wide subbands, and each subband is further subdivided into 64 12.2 kHz wide channels. For a more complete description of the Apertif system see \cite{vanCappellen_2021}. Apertif was designed to be a survey instrument and operated in survey mode between the 1st of July 2019 and 28th of February 2022. Apertif carried out a two tiered imaging survey, with a shallow, wide-area and a medium-deep, small-area component (see Hess et al. in prep), and a time domain survey (see \citealt{Maan_2017, vanLeeuwen_2022}). All of the Apertif surveys are legacy surveys and are providing publicly available data products. The first data release of Apertif (DR1) was in November 2020 (\citealt{Adams_2022})\footnote{See also: \url{http://hdl.handle.net/21.12136/B014022C-978B-40F6-96C6-1A3B1F4A3DB0}}. 

Accurate knowledge of the primary beam of a telescope is important for the correct deconvolution, for primary beam correction and for mosaicing of radio images. It is furthermore essential for the accurate instantaneous localisation of transients such as Fast Radio Bursts (\citealt{Connor_2020, vanLeeuwen_2022}). An incorrect primary beam can lead to severe imaging artefacts and an incorrect flux scale. This can especially be a problem for telescopes with PAF receivers, where the numerous simultaneously formed primary beams can all have significantly different shapes in contrast to telescopes with traditional receivers with only one primary beam shape. An example for imaging artefacts due to an inaccurate primary beam model are concentric ring shaped direction dependent errors (DDEs) around bright continuum sources. Figure 3 in \cite{Adebahr_2022} shows an example of such DDEs in the Apertif data. These DDEs can be mitigated with rigorous flagging of CBs that have low quality data (e.g. if CBs on one antenna have strong distortions compared to the other antennas, see Figure~\ref{fig:elements}) and advanced imaging techniques that include direction dependent calibration or custom antenna based primary beam models. This makes it essential to map and understand the shape of the primary beam as a function of frequency and time. As part of the Apertif imaging surveys data products we provide primary beam maps for each of the 40 digitally formed CBs. 

In this paper we present the characterisation of the Apertif CBs using two independent methods: drift scan measurements, and a Gaussian process (GP) regression method. In Section~\ref{section:CBs} we shortly describe how the CBs are formed. In Section~\ref{section:Methods} we describe the drift scan observations and the method used to create the beam maps. In Section~\ref{section:beam_models} we present the properties of the CBs, such as beam sizes, frequency dependence, time dependence, and a comparison between the drift scan beam maps and the GP beam maps. Finally in Section~\ref{section:Conclusions} we summarise our findings.

\section{Apertif compound beams}
\label{section:CBs}

The Apertif CBs are formed by applying beam weights to the data from the individual antenna elements. This can be done with different beam forming optimisation methods (e.g. \citealt{Jeffs2008, Ivashina2011}). Beam forming can be optimised for signal to noise or for the shape of the CBs. When maximising signal to noise the shape of the CBs gets gradually distorted with distance from the optical axis of the focal plane. Apertif uses maximum signal to noise beam weights, which means that CBs away from the optical axis suffer coma distortions. Similar distortions are also observed for the CBs of ASKAP (e.g. \citealt{McConnell2016}).

\begin{figure}
  \centering
  \includegraphics[width=4.4cm]{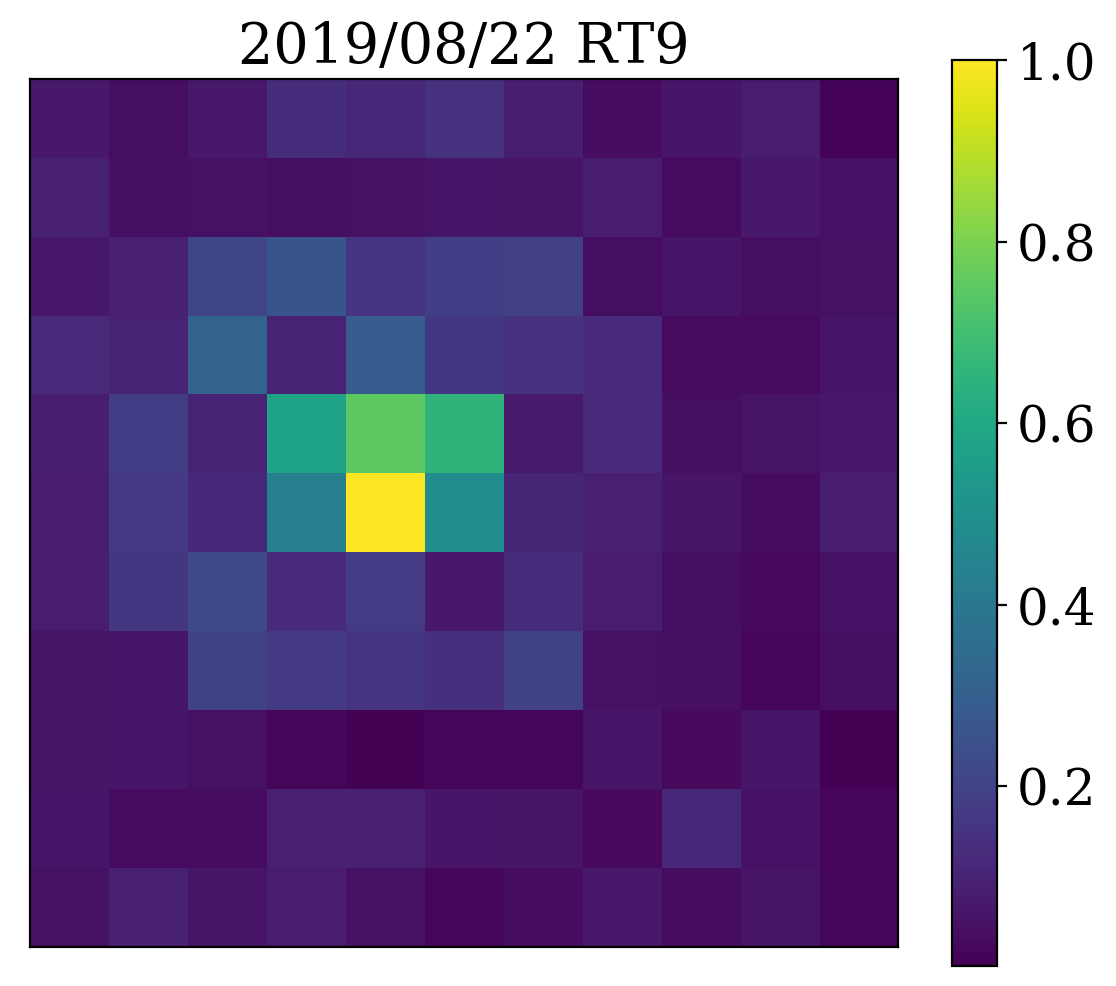}
  \includegraphics[width=3.8cm]{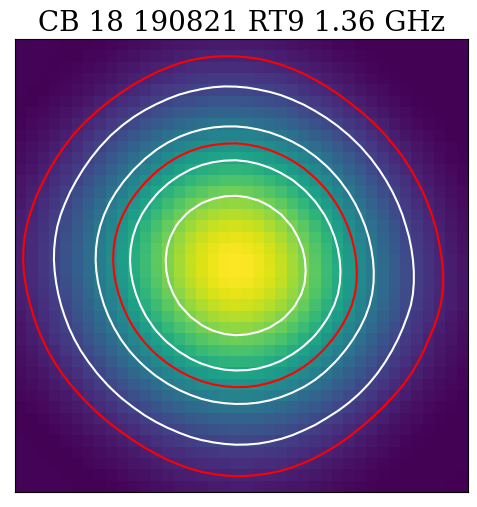}
  \includegraphics[width=4.4cm]{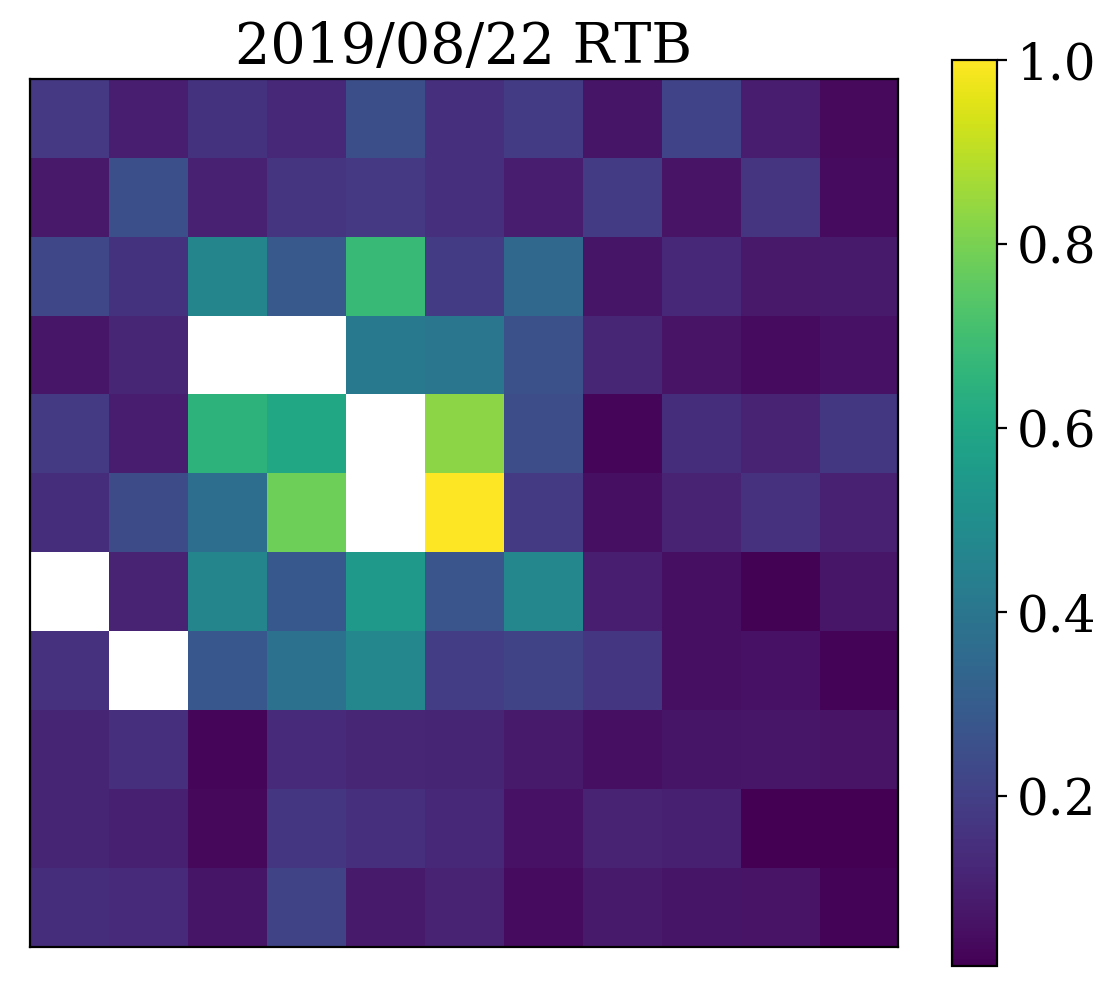}
  \includegraphics[width=3.8cm]{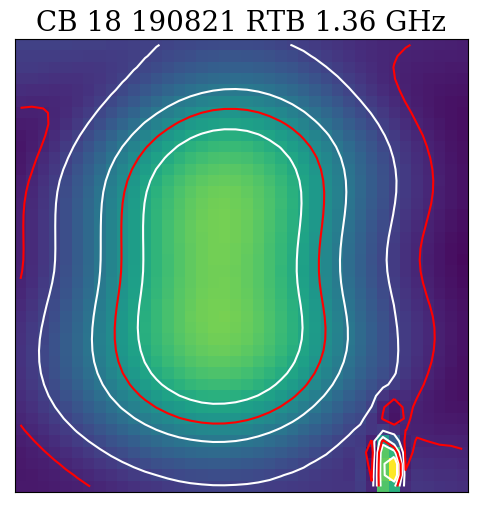}
  \caption{Left: Beam weights for Beam 18 on 2019/08/22 for antenna RT8 (top) and RTB (bottom). Each pixel corresponds to an antenna element on the PAF. Broken elements are white pixels in the image. Right: the corresponding CB map from a drift scan measurement. White contour levels are: 0.2, 0.4, 0.6, 0.8; red contour levels are: 0.1 and 0.5. The colour scale is the same for all four images. Most of the signal for a given beam is contributed by 9 antenna elements. If one of these antenna elements is broken the CB for the given antenna will be distorted. In this case 2 of the highest contributing elements on antenna RTB were broken.}
              \label{fig:elements}%
    \end{figure}

In the case of Apertif, CBs are formed so that most of the signal is contributed by only a few elements ($\sim$ 9 elements) and hence the sensitivity and the shape of the CBs strongly depend on the properties and health of the main contributing elements. The health of the Apertif system is assessed through monitoring all hardware and software components of the telescope (see detailed description in \cite{vanCappellen_2021} Sections 9 and 11). The components of the system health that are most critical for the quality of the science observations are the state of the individual Vivaldi antenna elements and the signal path from them to the correlator. To monitor this, the power against frequency is recorded from all antenna elements. Occasionally the power drops from the nominal 94 dB to 70 dB for an individual element. When this happens, the element is considered malfunctioning and the signal path from it is disabled. This means, that broken elements are excluded from the beam weights measurements and the scientific observations. Depending on how many and which elements fail on a certain PAF some CB shapes can be significantly distorted. Figure~\ref{fig:elements} illustrates this by comparing the beam weights and beam maps of a PAF with broken elements to a PAF with no broken elements. The top row of Figure~\ref{fig:elements} shows the beam weights and the drift scan map of CB18 on antenna RT8 and the bottom row shows the same data on antenna RTB, where there are several broken elements. On the healthy PAF the main contribution for a beam comes from approximately nine elements and the resulting CB shape is regular, while the broken elements on RTB distort the CB shape. To mitigate these distortions, it is possible to repair the broken elements on the PAFs, however due to the complicated nature of the PAF maintenance, elements are only repaired once a critical number (5 to 7 elements) fail on a PAF. On the upside, because of the small number of elements contributing to each individual CB the correlated noise between CBs is relatively low, on the level of a few percent.

CBs are formed for each 781.25 kHz wide subband and the beam characteristics can change across the subbands. A particular problem can be strong intermittent radio frequency interference (RFI) present in some subbands during the beam weight observations. This issue has been improved since December 2019; subbands strongly affected by RFI are since flagged more efficiently and beam weight values are interpolated from the nearby RFI free subbands. Beam weights for Apertif are measured at the start of every observing run, which is typically once in every two weeks. With every new beam weight measurement the properties of the CBs can change depending on the health of the antenna elements, the health of the system and the RFI environment during the beam weight measurement.

\section{Deriving CB maps}
\label{section:Methods}

To measure the shape of the response for each of the CBs we use two independent methods. One of these methods is performing drift scans on a bright continuum source, and reconstructing the beam shapes from the measured autocorrelations. The other method is comparing the flux of continuum point sources in the Apertif survey observations with the measured flux in an already existing, well calibrated catalogue, such as the NRAO VLA Sky Survey (NVSS; \citealt{Condon1998}), and constructing the beam shape with a Gaussian Processes regression method (see Section \ref{section:GP_maps}). 

During the drift scan measurement the PAF is at a fixed position on the sky and the observed source drifts through the field of view in a straight line. With the equatorial mount of the WSRT drift scans can only be performed in one direction along the R.A. axis. The separation between the individual drifts is 0.1 deg in declination. To scan the whole field of view of the 40 Apertif CBs - up to the 10 \% level of all the edge beams - 33 individual drift scans are needed, which together form a full set of drift scans \footnote{Drift scan observations are scheduled using the aperdrift code: \url{https://github.com/kmhess/aperdrift}, \url{https://doi.org/10.5281/zenodo.6545764}}. Figure~\ref{fig:driftscan} illustrates this process, where the numbered grey circles represent the CBs with a diameter of 0.6 deg, and the blue lines represent the individual drift scans. The start of each drift scan is marked by a grey triangle. 
    
\begin{figure*}
  \centering
  \includegraphics[width=8.8cm]{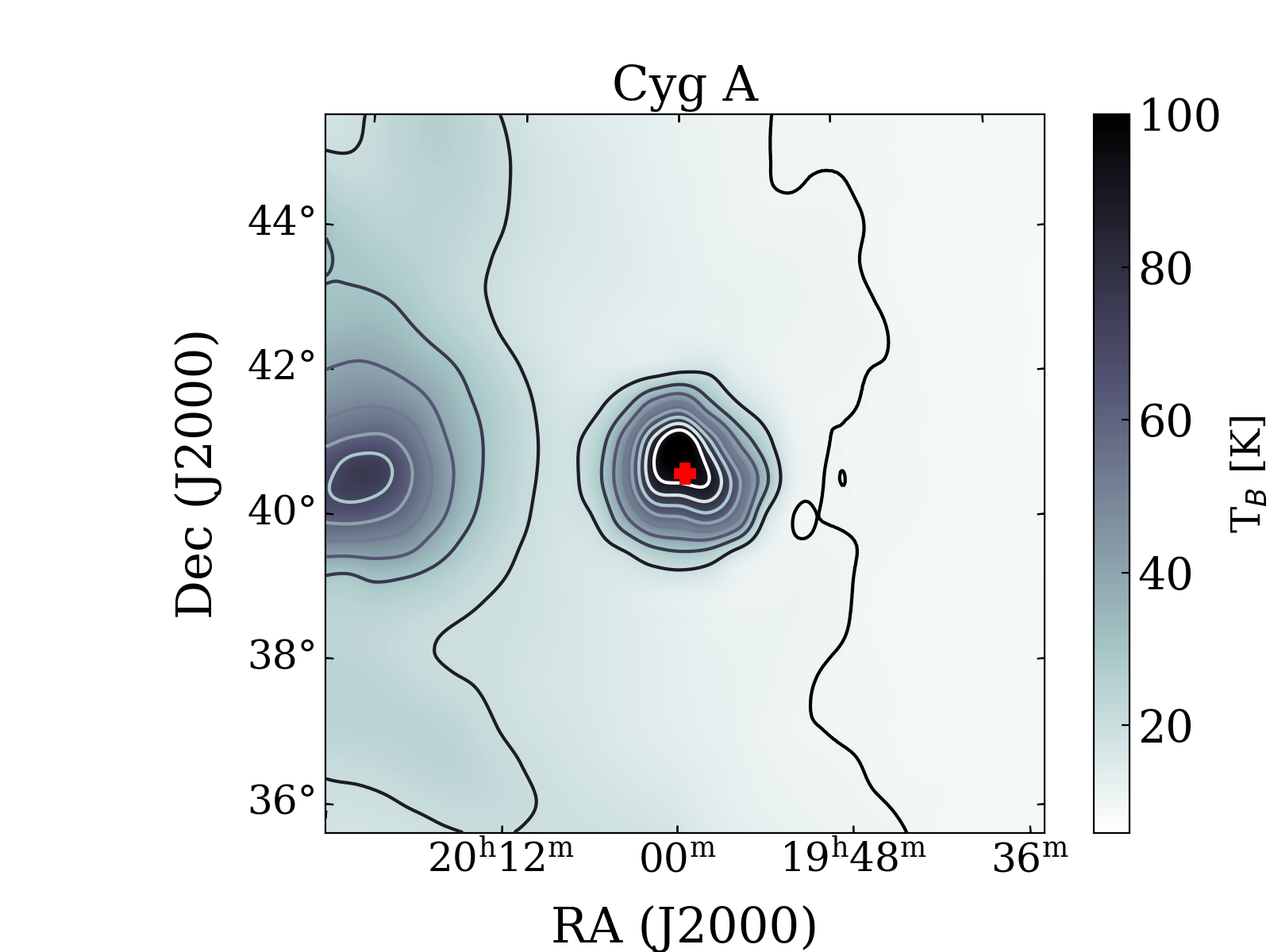}
  \includegraphics[width=8.8cm]{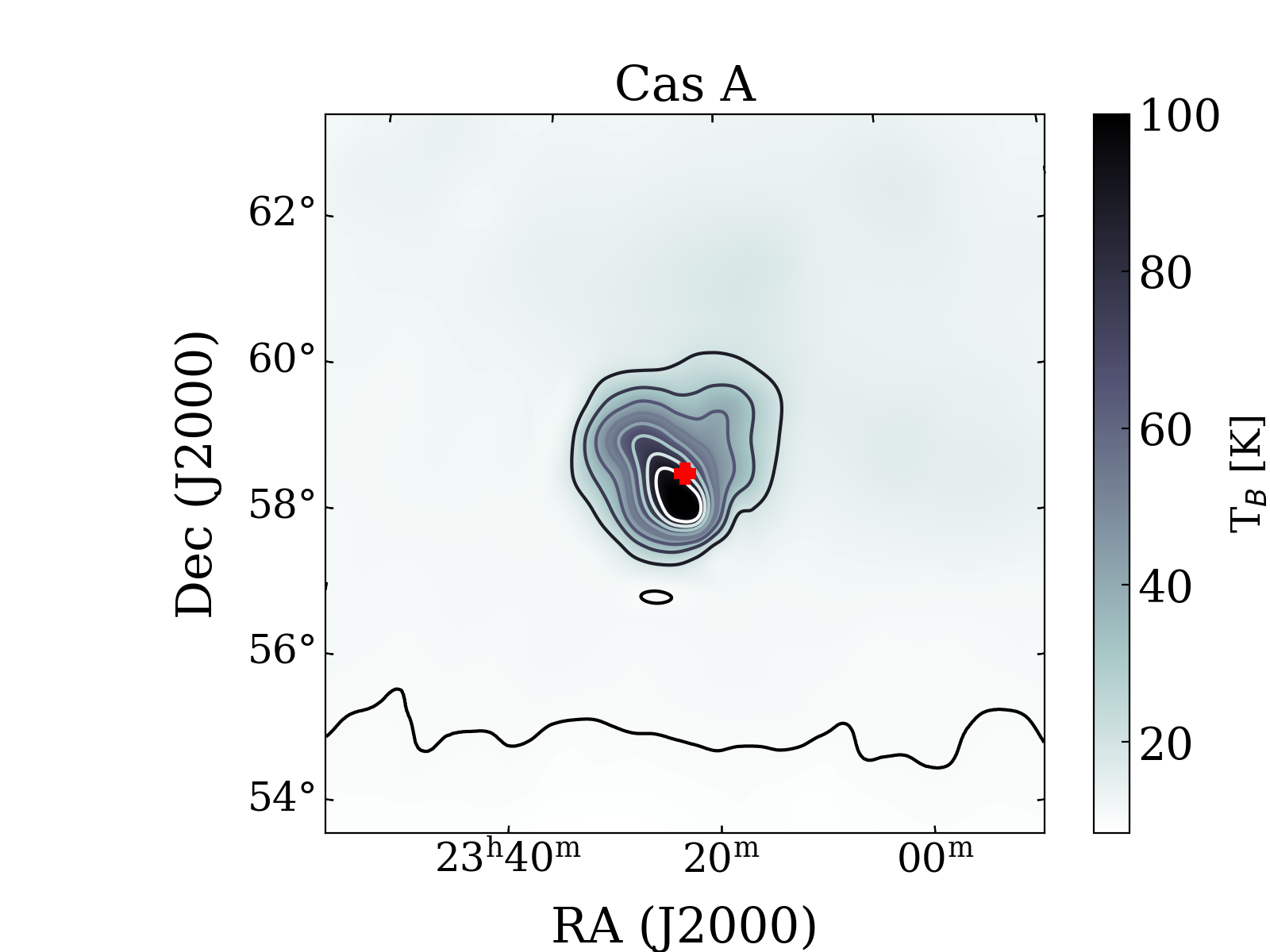}
  \caption{Continuum map of Cyg A and Cas A from the Dwingeloo 820 MHz survey \citep{Berkhuijsen1972}. Both images are 10 $\times$ 10$^{\circ}$ with a spatial resolution of 1.2$^{\circ}$. Contour levels are 10, 20, 30, 40, 50, 60, 70, 80, 90 K. }
              \label{fig:sources}%
    \end{figure*}

We perform drift scan measurements periodically on a compact, bright radio source: Cygnus A (CygA) or Cassiopeia A (CasA). CygA has an integrated flux of 1589 Jy \citep{Birzan2004} at 1.4 GHz, Cas A has an integrated flux of 2204 Jy \citep{DeLaney2014} at 1.4 GHz. Both of these sources are marginally resolved by a single WSRT dish, however they are the only sufficiently bright and compact sources available to perform drift scan measurements. Figure~\ref{fig:sources} shows the continuum images of Cyg A and Cas A from the Dwingeloo 820 MHz survey\footnote{\url{http://www3.mpifr-bonn.mpg.de/survey.html}} \citep{Berkhuijsen1972}. The Dwingeloo telescope has the same diameter (25 m) as the WSRT dishes, hence the angular resolution is comparable to the Apertif auto correlation data. For constructing the beam maps we use the auto correlation data from the telescope. The cross correlation data can not be used to measure the CB shapes because the fringe rotation of Apertif is faster for the long baselines than our time sampling (10 second). This means that we only detect signal on the short baselines but not on the long baselines during the drift scan measurements. Decreasing the time sampling for the observations is not possible in the imaging mode of Apertif. In addition, a finer time sampling would be impractical considering the already large data volume of the drift scans. A full set of drift scans on Cyg A is $\sim$15.5 TB and a full set on Cas A is $\sim$22.5 TB. Since we only use the auto correlations for measuring the CB shapes, we reduce the data size to 3.1 TB and 4.5 TB for the Apertif Long Term Archive by deleting the cross correlations from the drift scan measurement sets. The archived drift scan auto correlation data are available to the community upon request.  

We perform drift scan measurements approximately once per month, because of their time consuming nature. One set of drift scans on Cyg A takes $\sim$14 hours and one set on Cas A takes $\sim$ 20 hours. The difference in observing time and data size is due to the different declination of the sources ($\delta_{CygA} = 40.73^{\circ}$, $\delta_{CasA} = 58.81^{\circ}$). Drift scans take longer on sources at higher declination. The relatively long observing time also means that a full set of drift scans needs to be observed in at least 2 separate observing windows, since the equatorial mount of the telescopes can only track sources for 12 hours in a day. 

\section{CB maps}
\label{section:beam_models}

For the first 19 months of the imaging survey the observations were conducted between 1130-1430 MHz. However, due to strong RFI only 150 MHz between 1280-1430 MHz were processed. From this processed band the first 30 MHz are also regularly flagged due to RFI. For the same reasons we only derive beam maps for the upper half of the band, for the frequency range 1300-1430 MHz. In January 2021 the Apertif imaging surveys observing frequency was updated to 1220-1520 MHz to better avoid RFI. Accordingly, for those observations we updated the CB map range to 1300-1520 MHz, the first 80 MHz of the bandwidth are continued to be the flagged due to RFI. 

\begin{figure*}
   \centering
   \includegraphics[width=16.4cm]{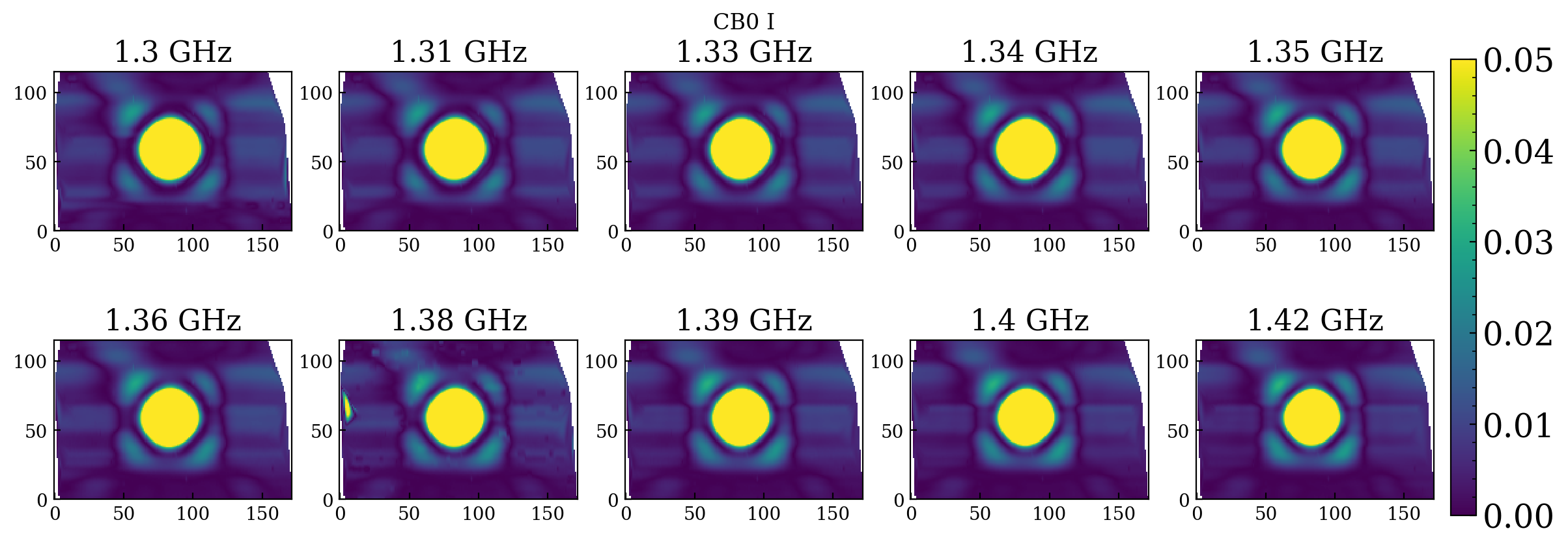}
   \includegraphics[width=16.4cm]{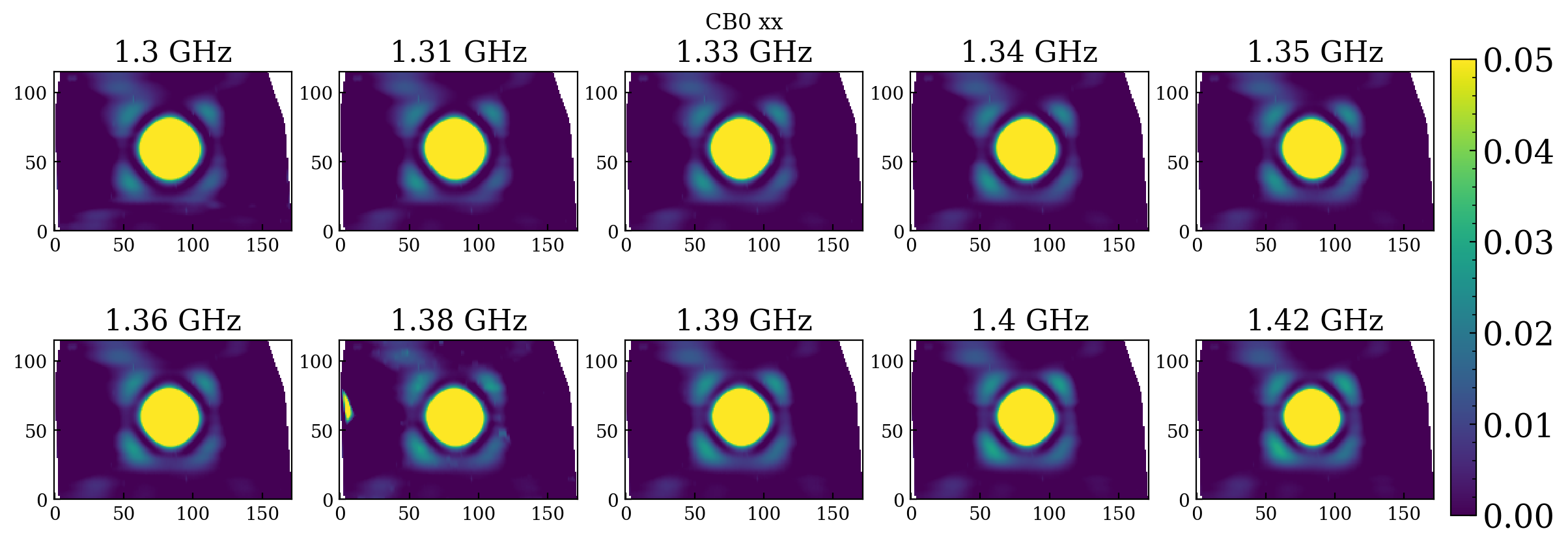}
   \includegraphics[width=16.4cm]{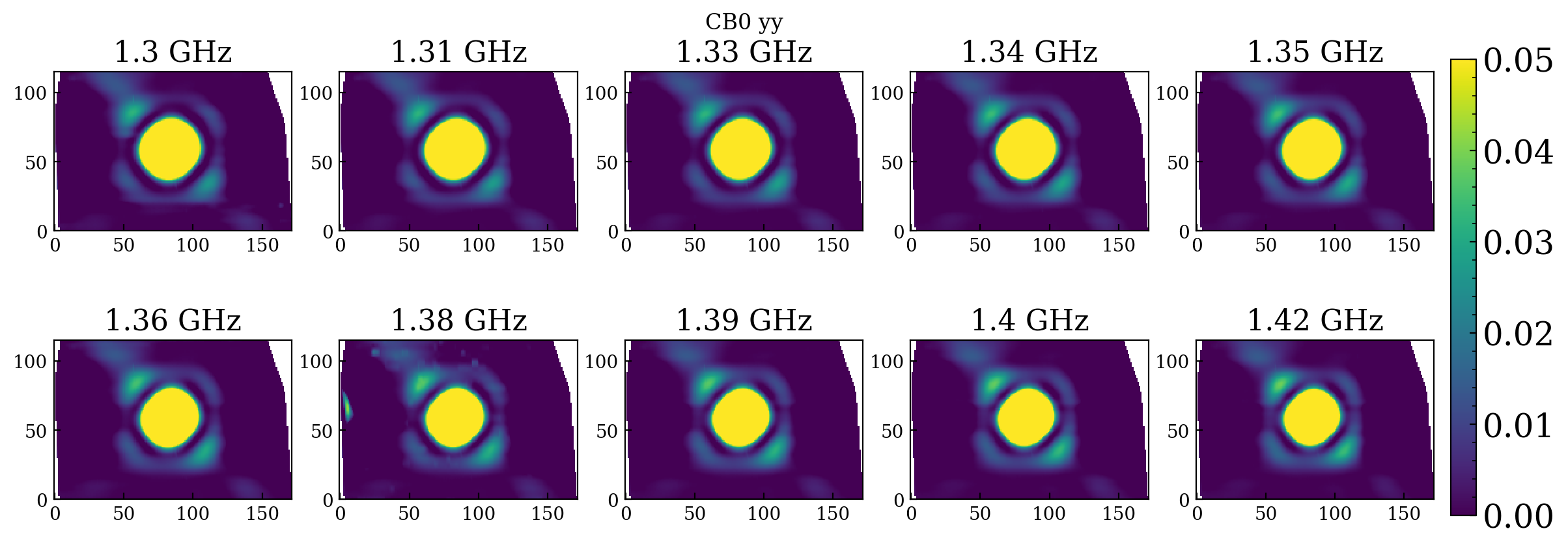}
   \includegraphics[width=16.4cm]{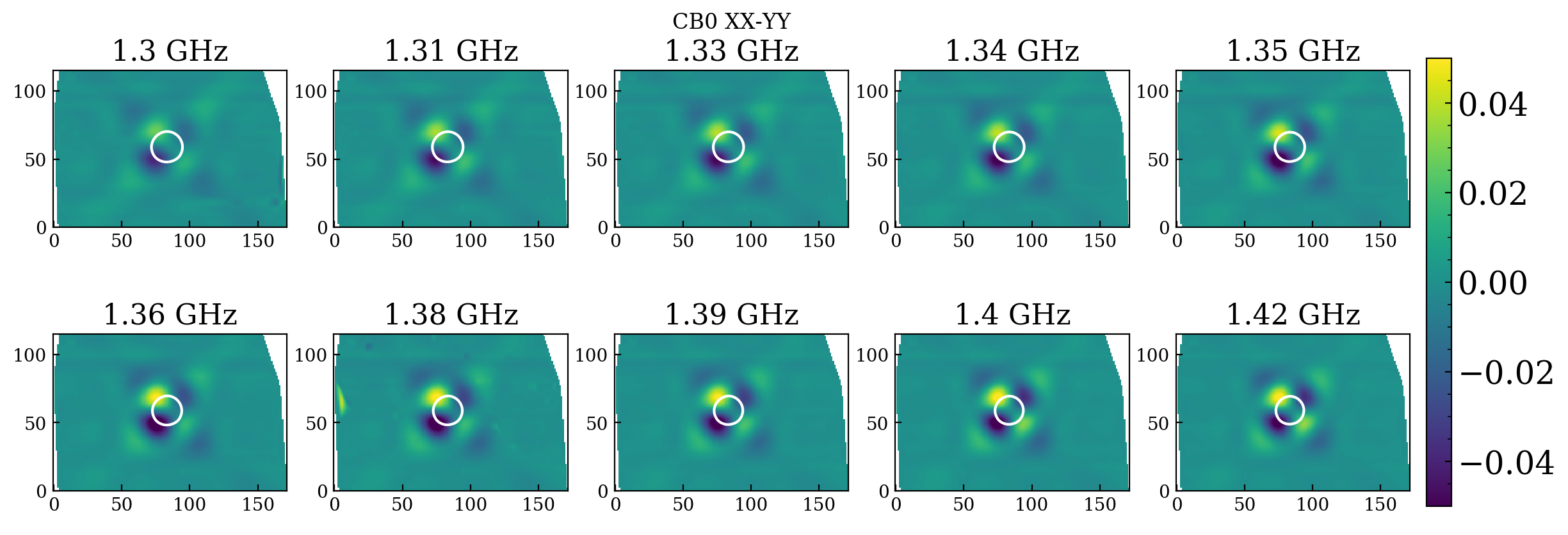}
   \caption{Beam maps for the 201028 data set CB 0 at all 10 frequency bins. From top to bottom: I, XX, YY, and XX-YY polarisation. The white circle indicates half power in the I polarisation. Each map is 3.36 x 2.3 deg.}
              \label{fig:beams_maps}%
    \end{figure*}

\begin{figure}
   \centering
   \includegraphics[width=8.4cm]{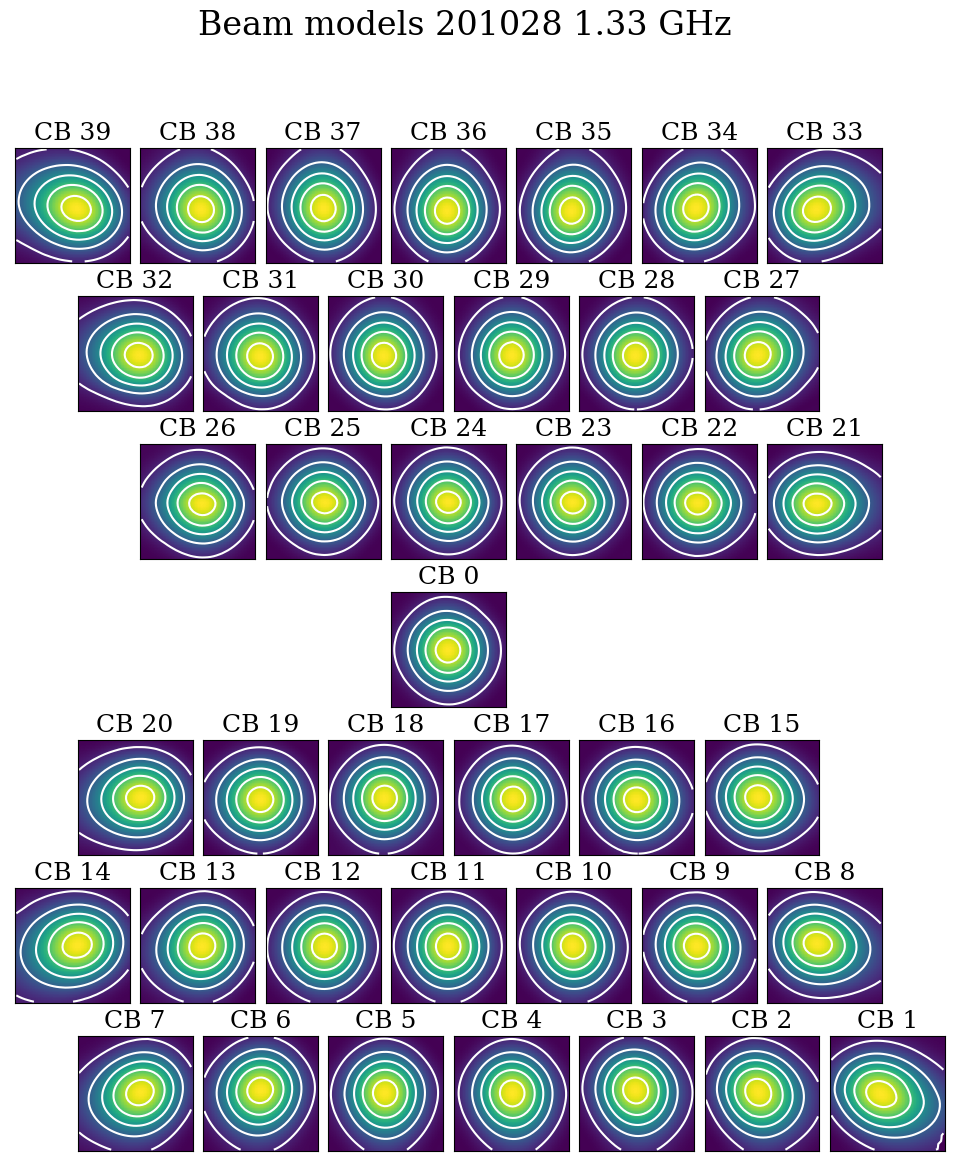}
   \caption{Smoothed beam maps for all 40 Apertif CBs for the 201028 data set at 1.33 GHz. Contour levels are: 0.1, 0.3, 0.5, 0.7, 0.9.}
              \label{fig:beams_chan5}%
    \end{figure}

We construct the CB maps from the auto correlations with the following steps:
\begin{enumerate}
    \item We extract the auto correlation data for each CB from each drift scan measurement set. We extract all measured polarisation data (XX, YY, XY, YX) in 10 frequency bins (1050 channels - 12.8 MHz - per bin), for all individual antennas, and averaged for all antennas as well. For data taken after January 2021, at the new observing frequency, we use 18 frequency bins (1000 channels - 12.2 MHz - per bin).  
    \item From the frequency binned auto correlation data we construct fits images for all 40 CBs. This step is done by combining the individual drift scans and gridding them onto a co-moving RA - Dec grid with the observed continuum source. For this we use the {\sc scipy} package: {\sc interpolate.griddata} with cubic interpolation. Maps are generated for XX, YY, I, and XX-YY polarisation and cover 3.36 $\times$ 2.3 deg (Figure~\ref{fig:beams_maps}).  
    \item Spline smoothing is applied to each CB map (to minimise the effect of RFI) and a smaller 1.1 $\times$ 1.1 deg (40 x 40 pixel) beam map, centred on each CB, is written into fits files (Figure~\ref{fig:beams_chan5}). These smaller CB maps are designed to use for primary beam correction and mosaicing the Apertif images and image cubes. They approximately cover the CBs to the 10\% level. The spline smoothing is done with the {\sc interpolate.RectBivariateSpline} module of {\sc scipy}, which performs a bivariate spline approximation over a rectangular mesh to smooth the data. We only produce these smaller beam maps for 9 or 17 frequency bins since the first frequency bin (at 1.3 GHz) is almost always strongly affected by RFI.
\end{enumerate}

The {\sc Python} code to produce the CB maps is publicly available on {\sc GitHub} \footnote{\url{https://github.com/apertif/aperPB}, \url{https://doi.org/10.5281/zenodo.6544109}}. To refer to specific drift scan data sets we use CB map IDs, which are based on the observing date of the drift scans in the format YYMMDD (e.g. 190821, 201028). In this work we use CB maps 190821, 190912, 200130, and 201028 for illustrations. The 190912, 200130, 201028 data sets are good overall representatives for most of the measured CB maps and the 190821 data set is chosen to illustrate the effects of broken PAF elements.

Figure~\ref{fig:beams_maps} shows an example of CB maps for the data set 201028 CB 0. We show CB 0 here, since it is the central CB of the Apertif footprint, directly along the optical axis of the dish \citep{vanCappellen_2021}, and is the most symmetric shaped CB. From top to bottom the figure shows I, XX, YY, and XX-YY polarisation CB maps for the 10 frequency bins. The spatial extent of the individual maps is 3.36 $\times$ 2.3 deg. The image is scaled to highlight the pattern of the side-lobes. The four fold symmetry of the CBs, particularly visible in the side-lobes, can be attributed to the support structure of the receiver. A somewhat similar beam shape was also measured for the previous MFFE 1.4 GHz receiver on the WSRT by \cite{Popping2008}. Figure~\ref{fig:beams_chan5} shows the 1.1 $\times$ 1.1 deg smoothed CB maps for the same data set (201028) for all beams at 1.36 GHz. The beams in the centre of the footprint are relatively symmetric while the edge beams are quite elongated due to coma distortion and fewer contributing antenna elements towards the edges of the PAF. 

\begin{figure}
   \centering
   \includegraphics[width=8.4cm]{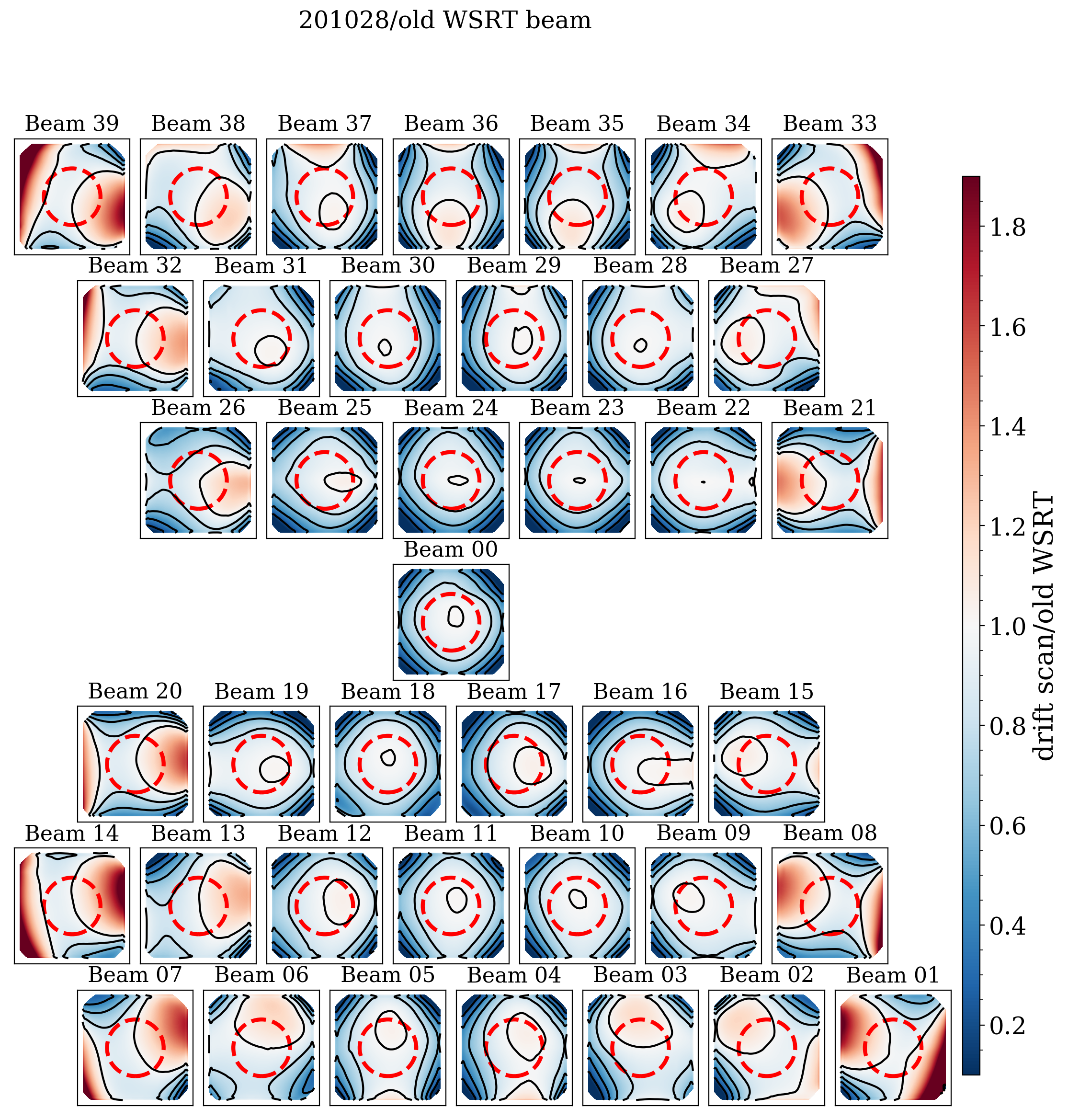}
   \caption{Comparing the CB shapes (201028 data set) to the old WSRT primary beam. The colours show the Apertif beam response divided by the previous analytic WSRT primary beam model. Contours are: 0.2, 0.4, 0.6, 0.8, 1.0. The red dashed circle marks 36 arcmin from the centre, which indicates the approximate FWHM for the CBs.}
              \label{fig:beam_maps_old_wsrt}%
    \end{figure}

We wish to emphasise that the use of the classic WSRT primary beam correction is not appropriate for Apertif. Figure~\ref{fig:beam_maps_old_wsrt} shows one set of measured CB maps (201028) divided by the previous analytic WSRT primary beam model (cos$^{6}(c\nu r)$, where $c$=68, $\nu$ is the observing frequency in GHz and $r$ is the distance from the pointing centre in degrees). In addition to the elongated shapes (and offsets) visible in the outer CBs, the Apertif CB value is generally smaller than the classic WSRT primary beam value. The Apertif PAF illuminates the reflector dish more uniformly than the old MFFE frontends, which leads to an increased aperture size and a smaller CB shape (see Figure 36 in \citealt{vanCappellen_2021}).

\begin{figure}
   \centering
   \includegraphics[width=8.4cm]{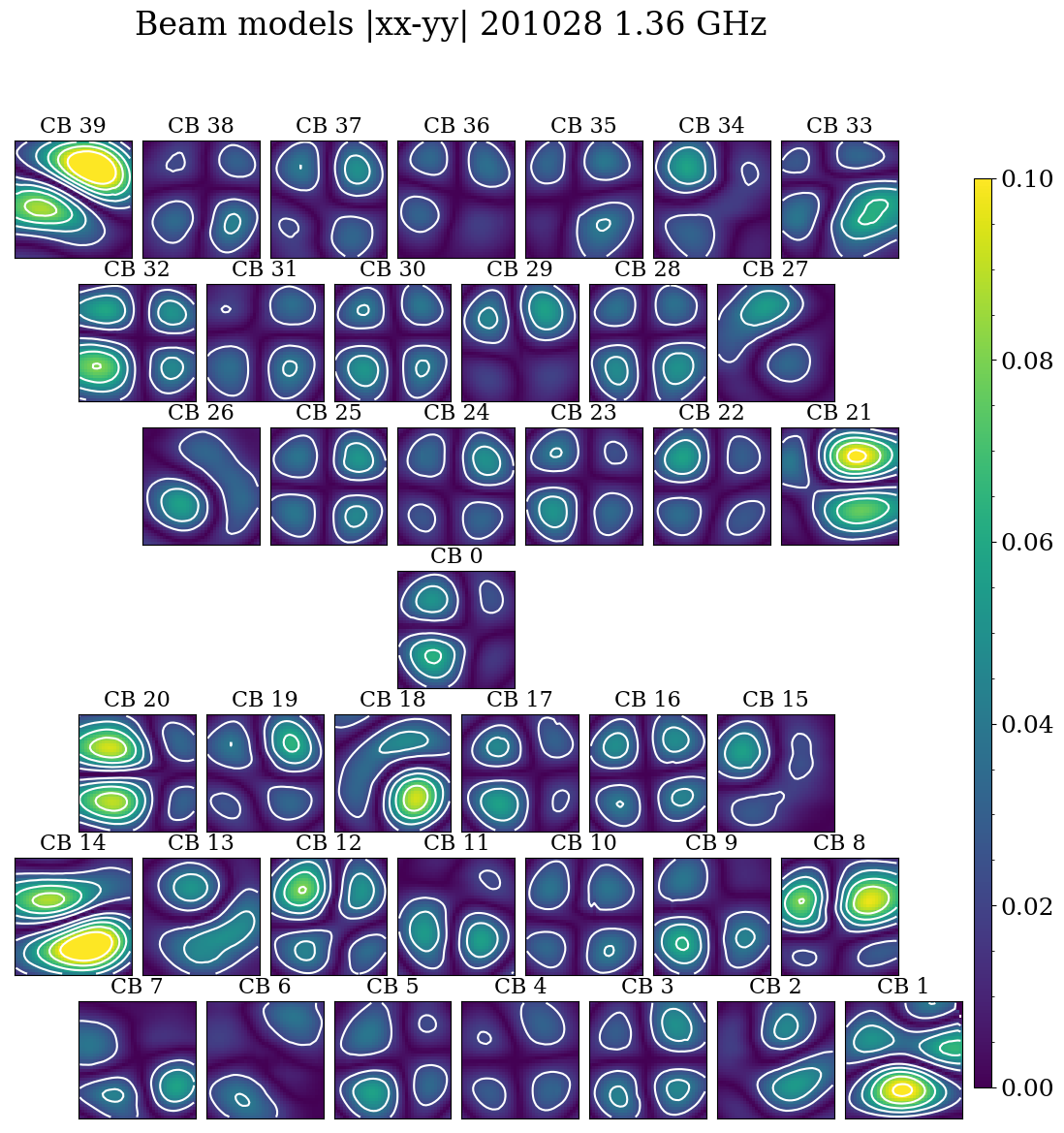}
   \caption{$\mid$XX-YY$\mid$ CB maps for 201028 to illustrate the polarisation leakage at 1.36 GHz. The colour scale is adjusted to show the most extreme polarisation behaviour for the CBs on the edge of the footprint. Contour levels are between 0 and 0.1 at an interval of 0.02.}
              \label{fig:beams_xx-yy}%
    \end{figure}

Drift scan maps can be used to evaluate some of the polarisation properties of the Apertif system. Figure~\ref{fig:beams_xx-yy} and the bottom pannel of Figure~\ref{fig:beams_maps} show the leakage properties of the system, where $\mid$XX-YY$\mid$ corresponds directly to the Stokes Q leakage. We find leakages on the order of 0.001 at the beam centres and 0.02 at at the 0.5 response level showing four-fold symmetry. For some of the edge beams, the leakage can be higher, up to 0.1 at the 0.5 response level of CBs 01 and 39, which have generally the most distorted CB shape. Analysing the polarisation measurements in the Apertif imaging observations we find that the fractional polarisation of polarised continuum sources stays constant at an approximate level of 1\% up to a CB response of 0.3. A more detailed analysis of the polarisation characteristics of the Apertif system is presented in \citealt{Adebahr_2022}.

\begin{figure}
   \centering
   \includegraphics[width=8.4cm]{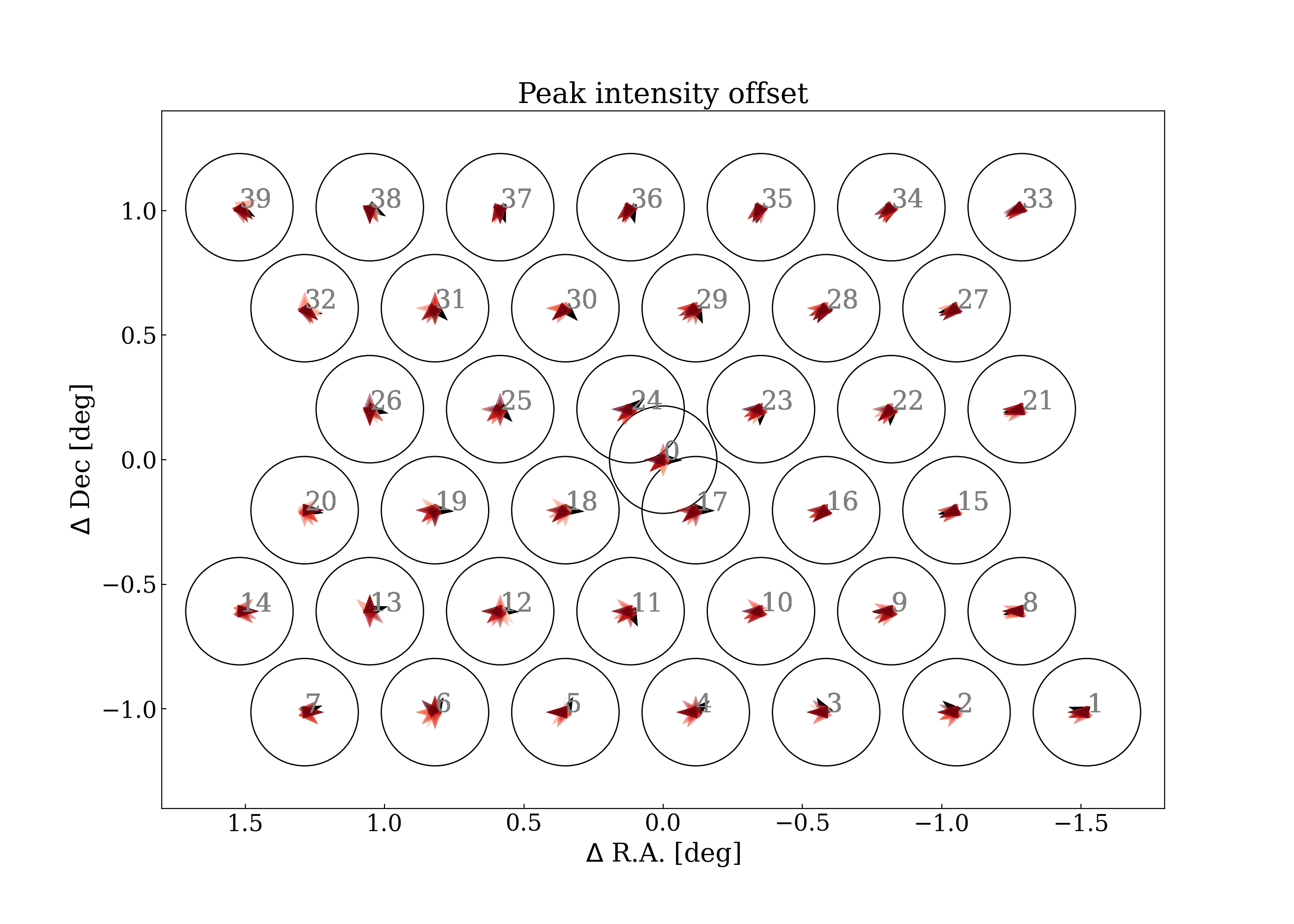}
   \caption{Field of view map of peak intensity offset from the centre for each compound beam. The arrowheads show the size (typically a few arcminutes) and the direction of the peak intensity offset. Black arrowheads represent the offset measured from the GP beams and red arrowheads the measurements from the individual drift scans.}
              \label{fig:centre_offset}%
    \end{figure}

There is a systematic variation in the position of the peak intensity of the CBs compared to the nominal centre of the beams. This offset is on average 1.5 arcminute, but can be up to 5 arcminutes for the edge beams. Figure~\ref{fig:centre_offset} shows the size and the direction of the peak intensity offsets from the nominal beam centre for all drift scan data sets in red and for the GP CBs in black. These offsets are intrinsic to the CBs and are present in the maps measured with both - GP and drift scan - methods. The offsets correspond to the elongated shapes of the CBs, i.e. the more elongated a CB is, the larger is the offset in the opposite direction of the elongation. This is also visible in Figure~\ref{fig:beams_maps_ant} where the nominal centre of the beam, indicated with a black dot, is offset from the most sensitive part of the beam, indicated with a red dot. The amplitude of the offset is stable for most CBs, however the direction of the offset does vary for some CBs. The CBs most affected by these variations are the ones in the left centre part of the footprint, which are also the CBs that have been the most affected by broken antenna elements during the Apertif survey observations. This also indicates that the CB shape variation over time is largely due to broken and repaired antenna elements.     

\subsection{CB maps for individual dishes}

\begin{figure*}
  \centering
  \includegraphics[width=5.2cm]{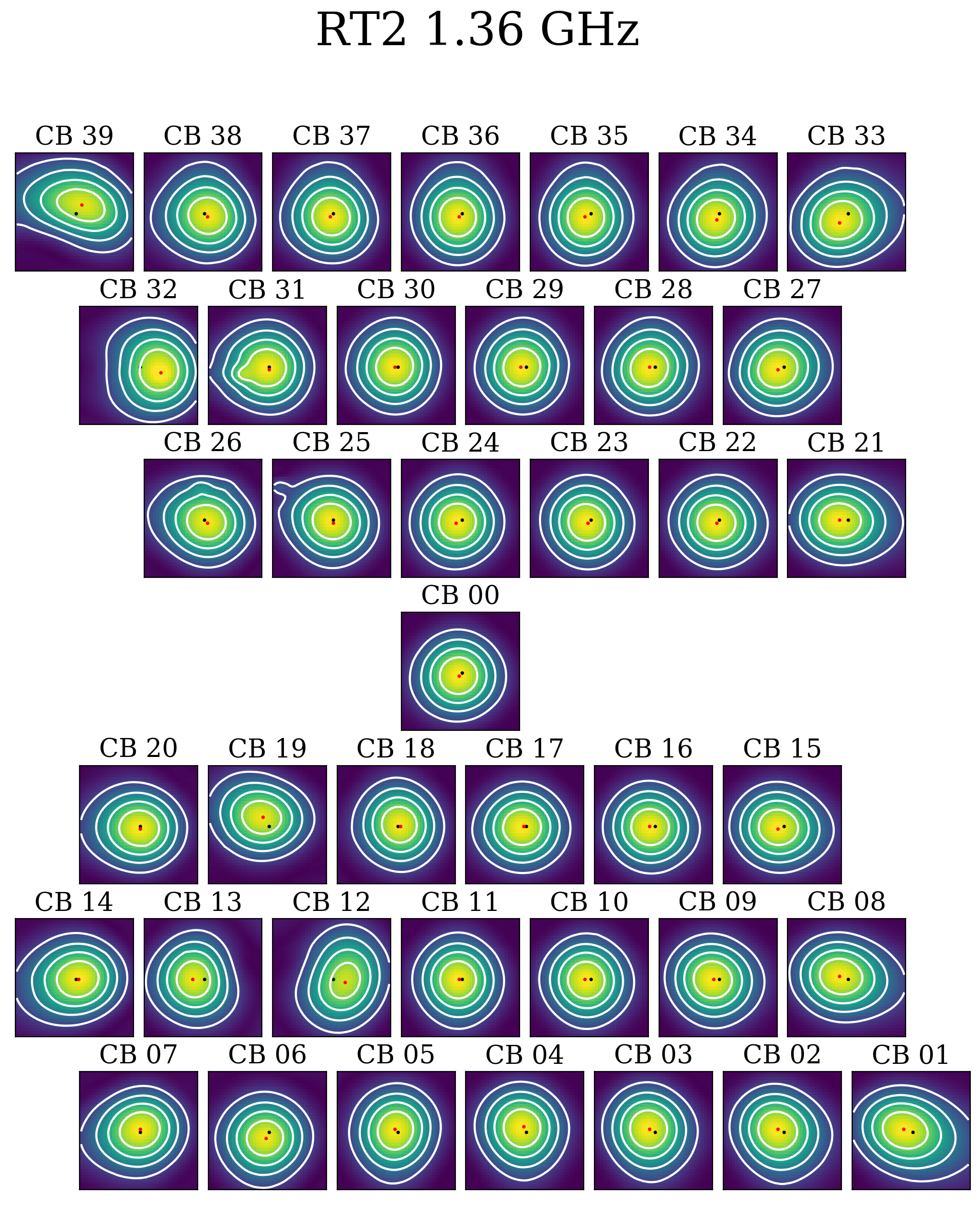}
  \includegraphics[width=5.2cm]{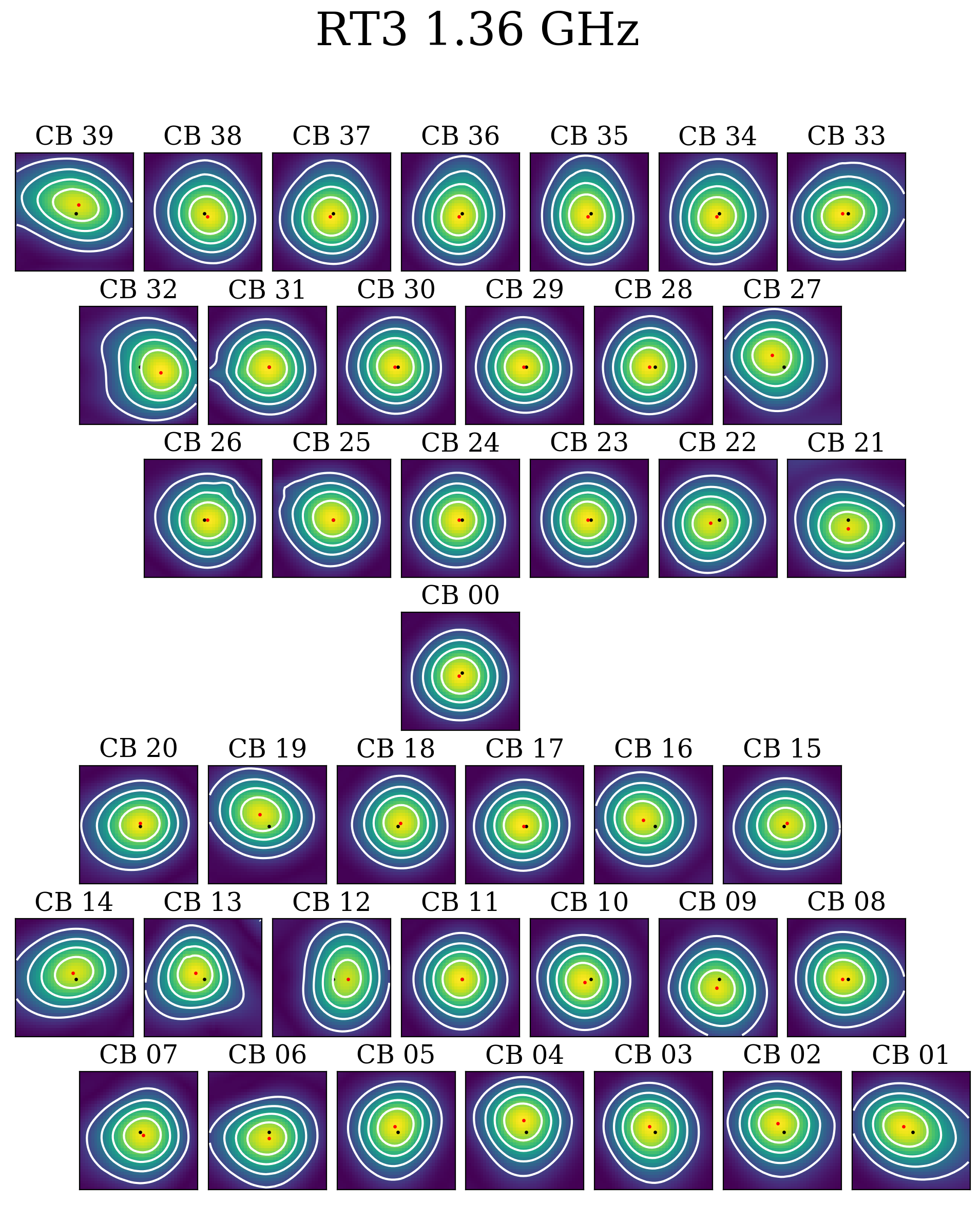}
  \includegraphics[width=5.2cm]{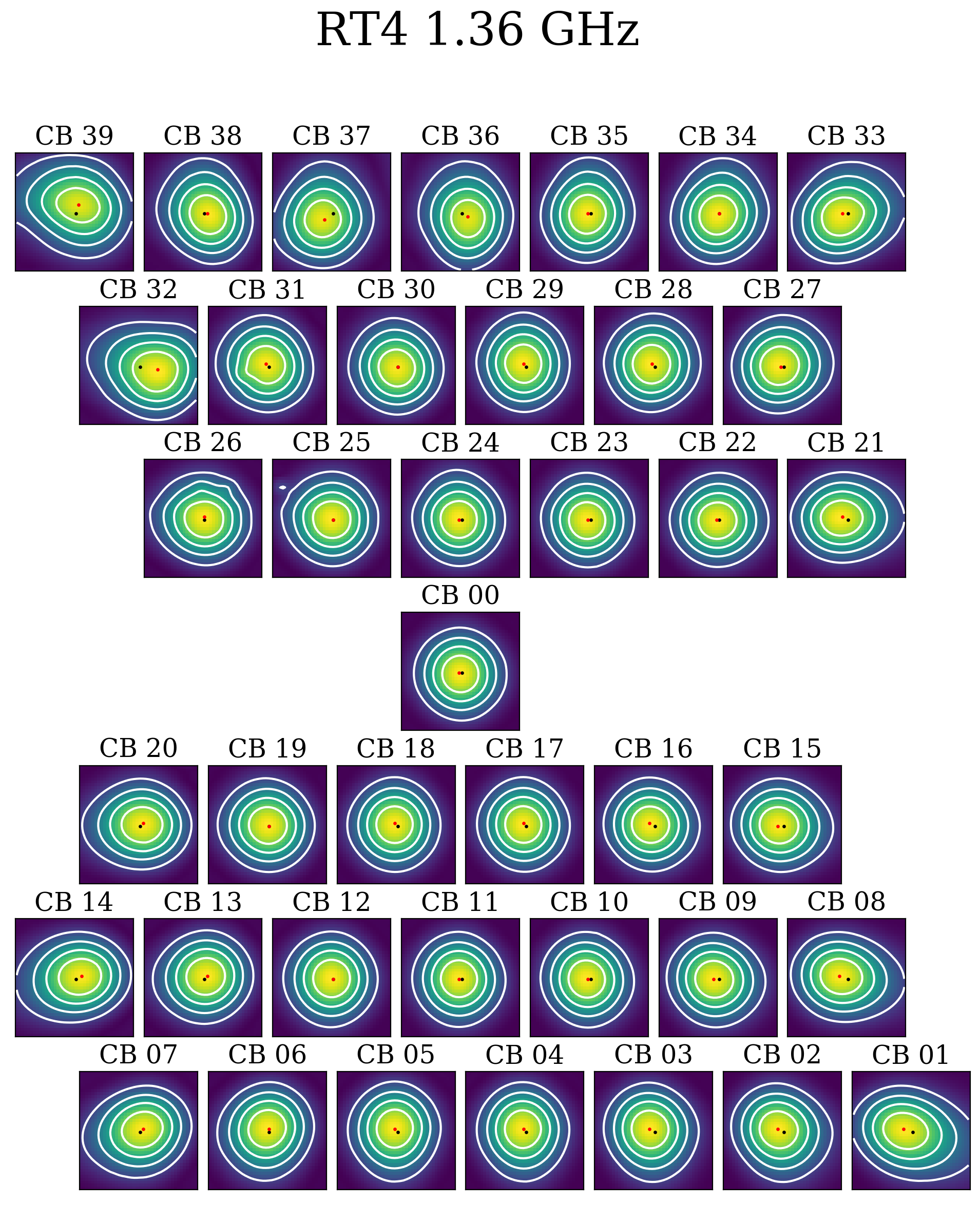}
  \includegraphics[width=5.2cm]{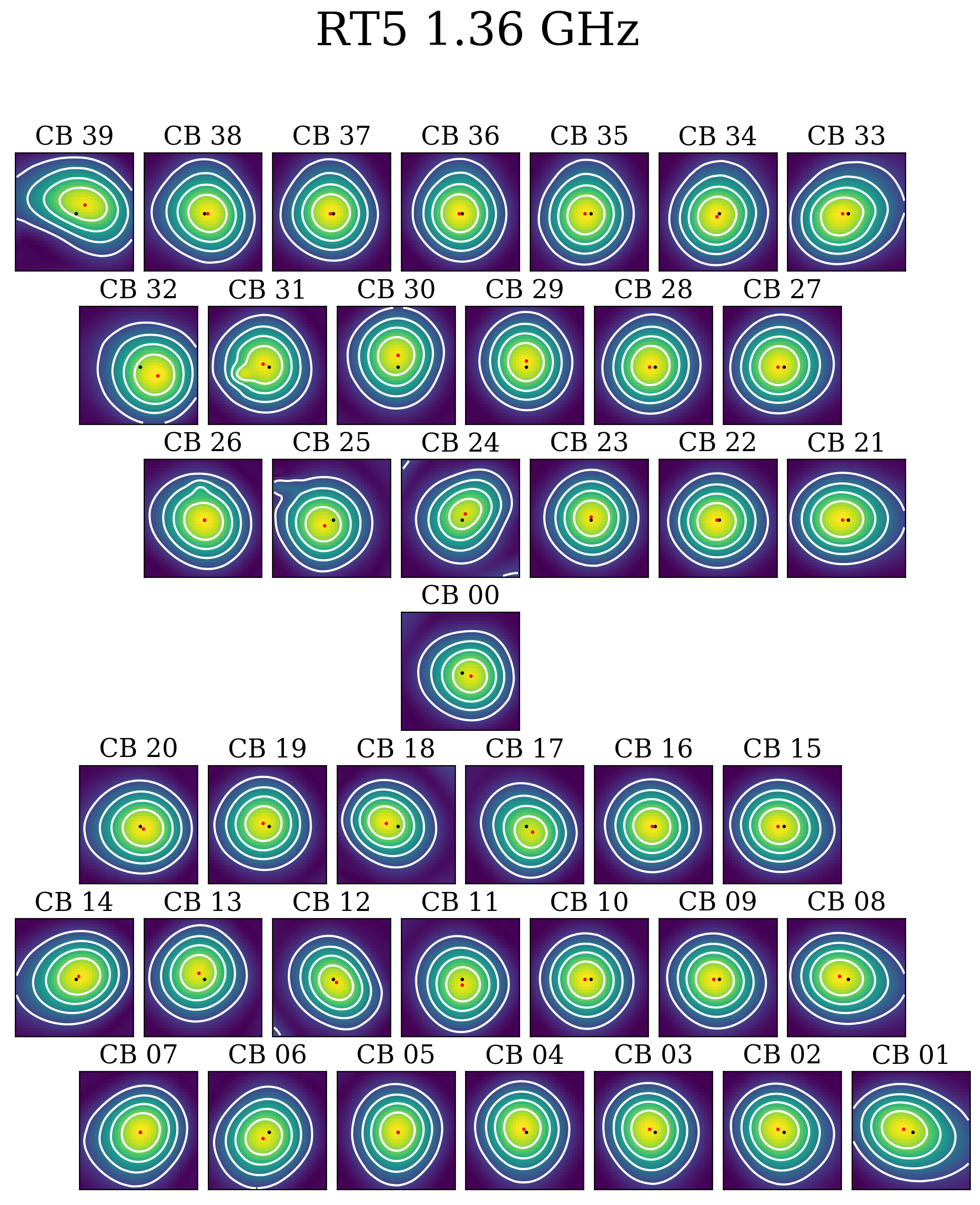}
  \includegraphics[width=5.2cm]{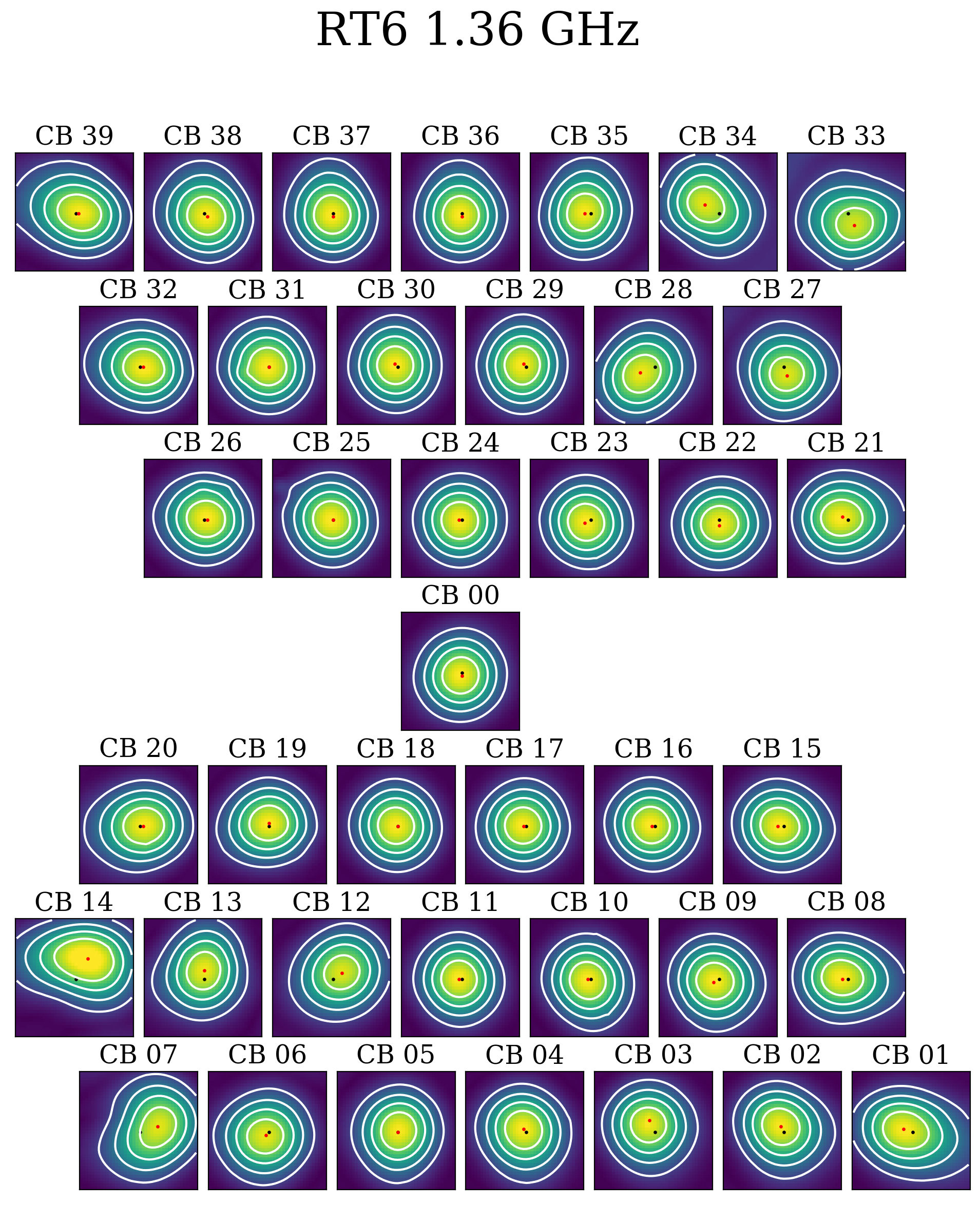}
  \includegraphics[width=5.2cm]{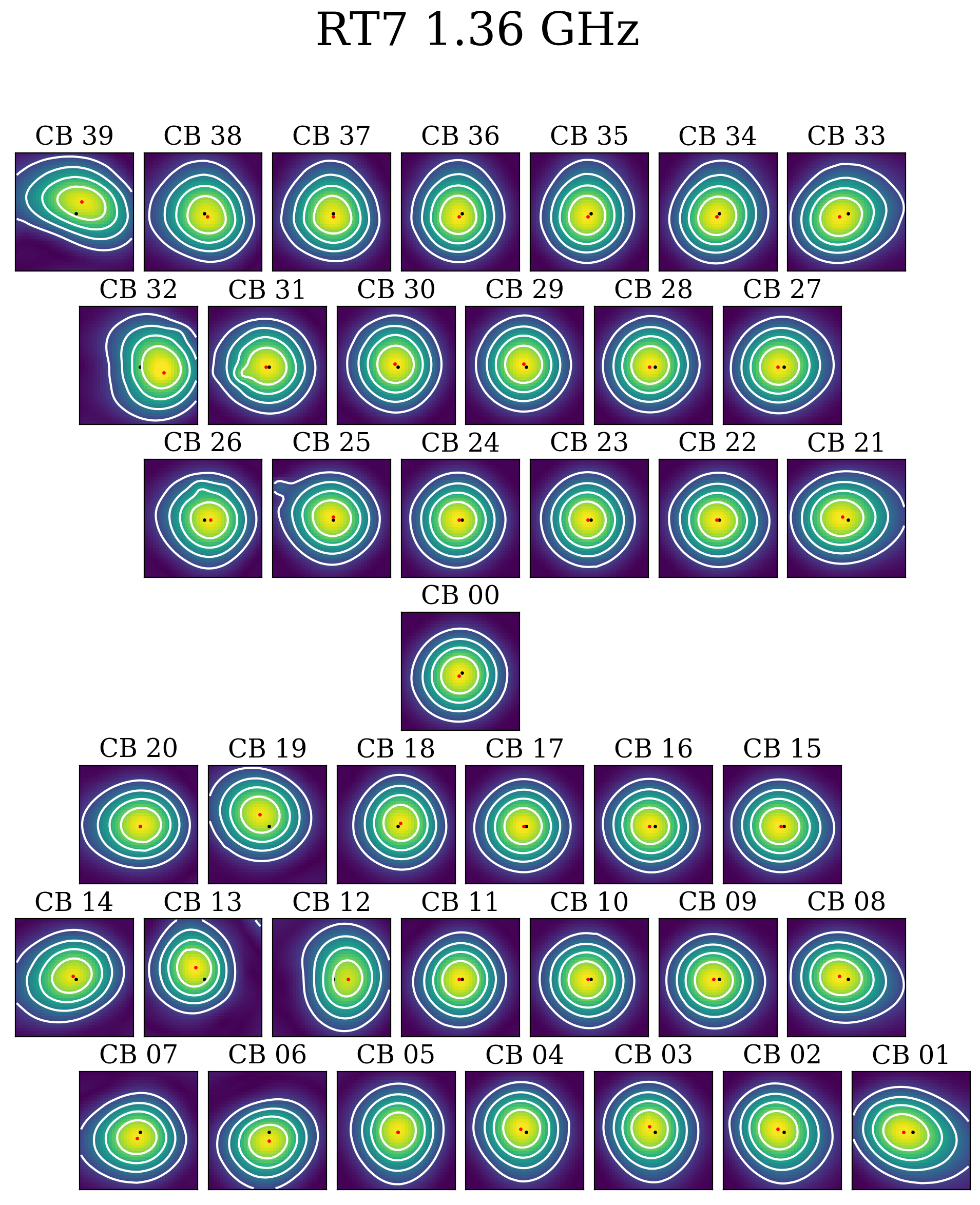}
  \includegraphics[width=5.2cm]{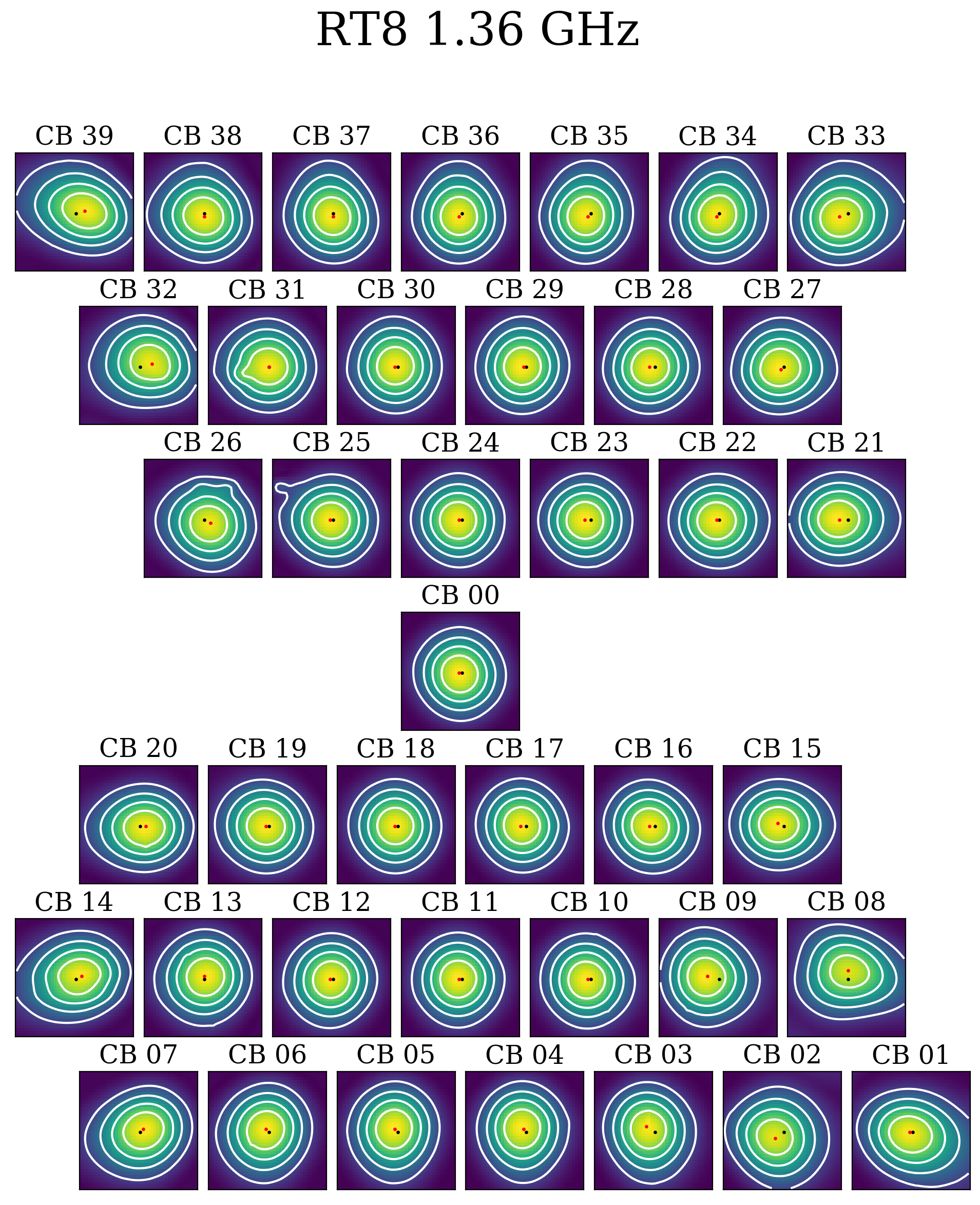}
  \includegraphics[width=5.2cm]{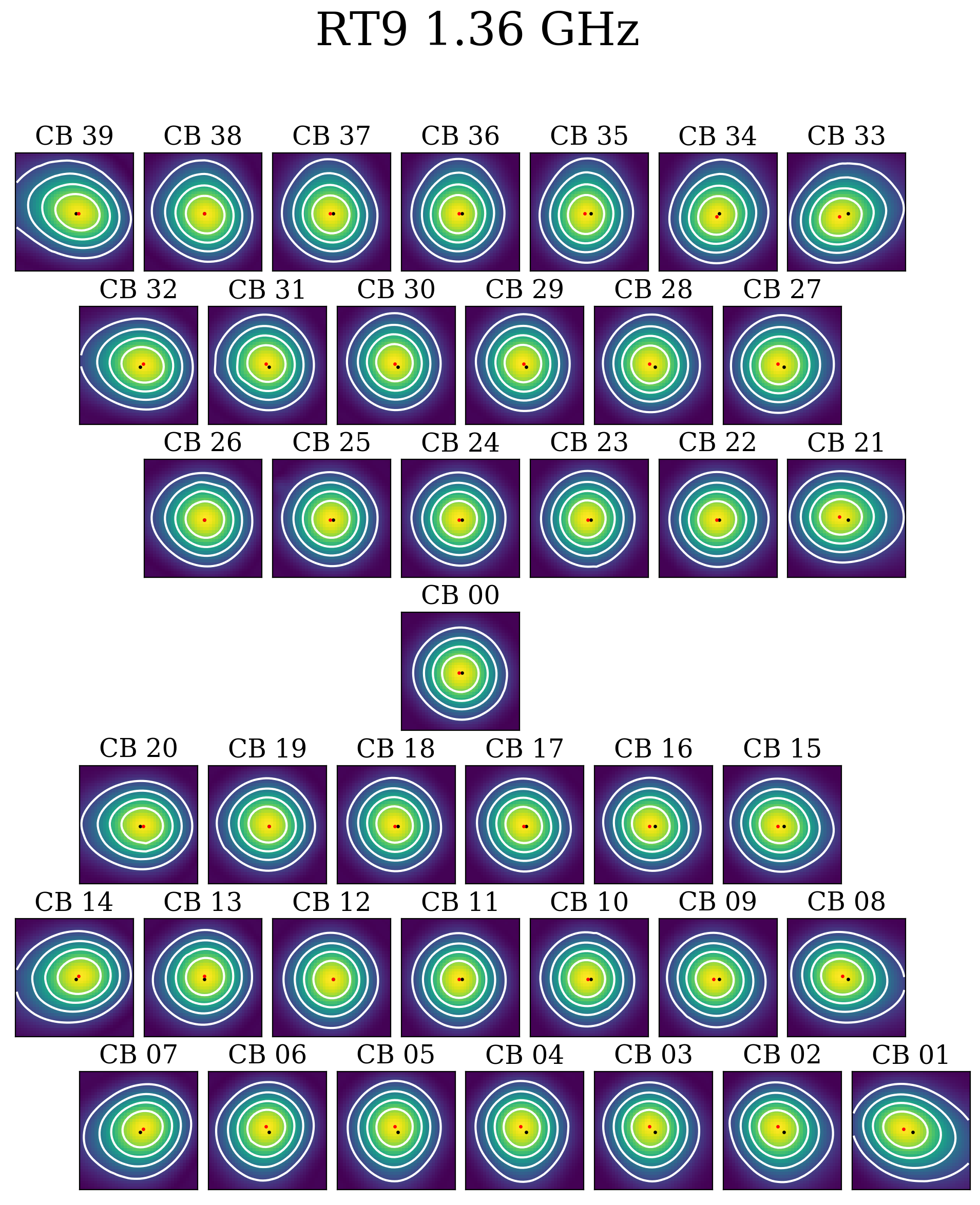}
  \includegraphics[width=5.2cm]{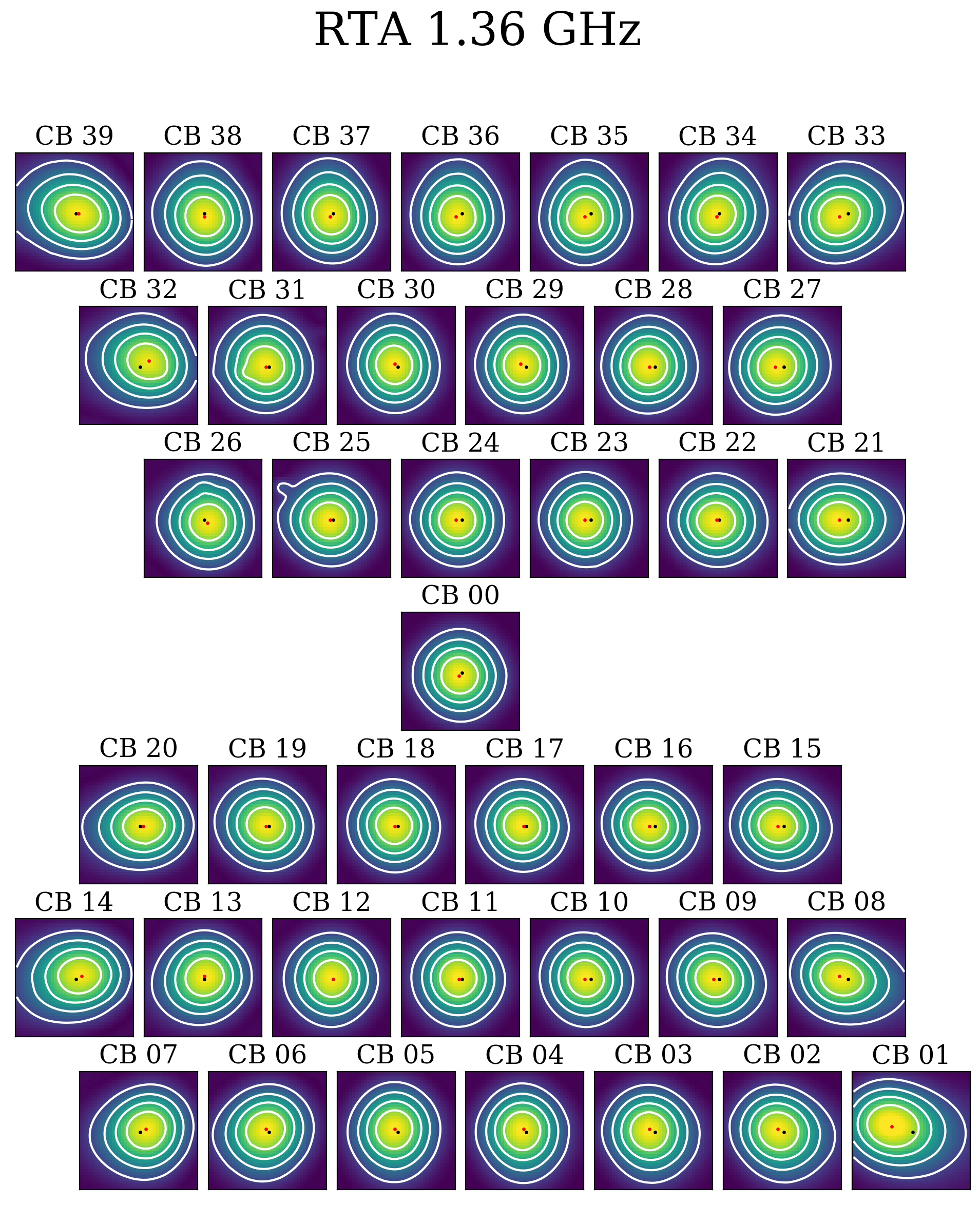}
  \includegraphics[width=5.2cm]{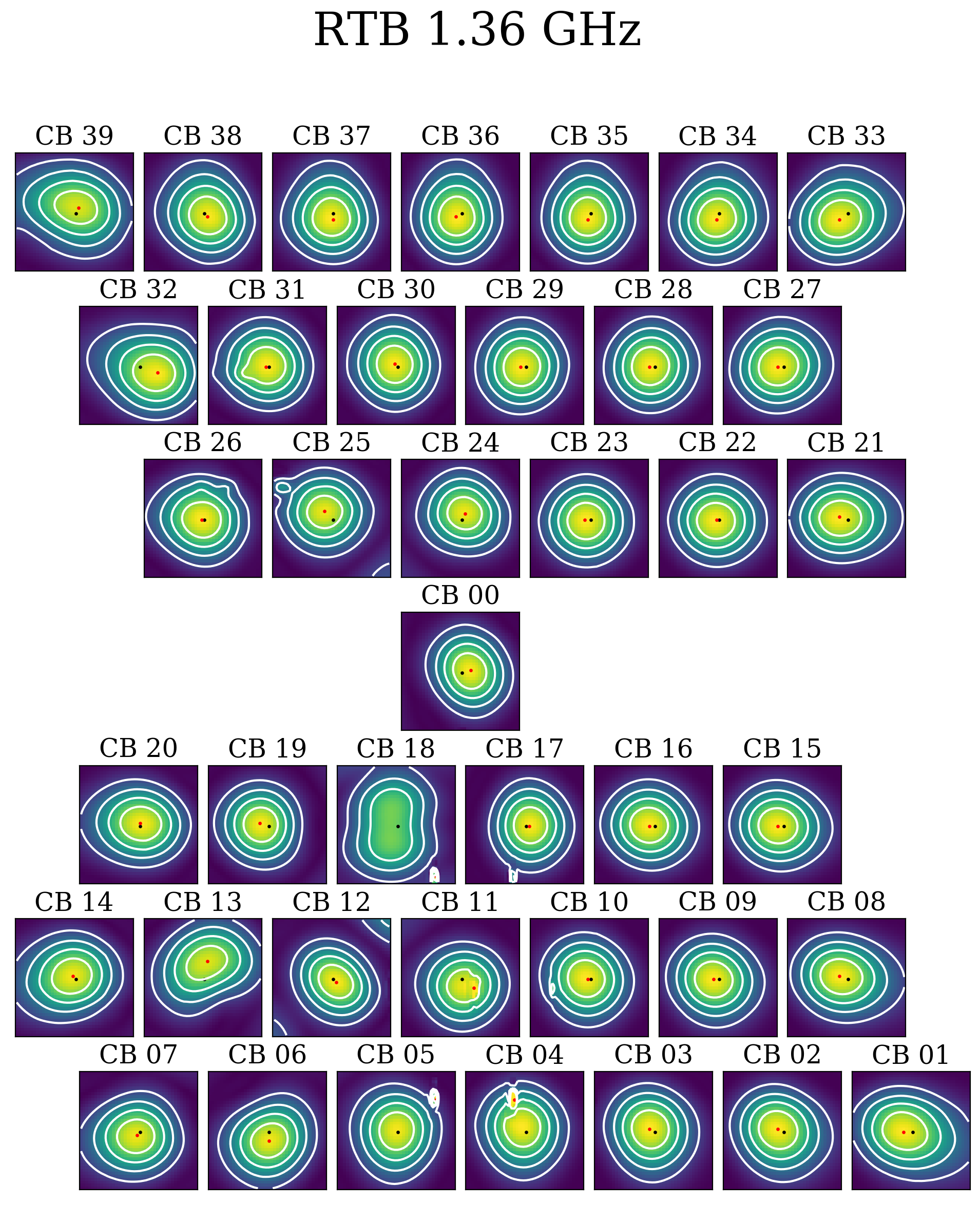}
  \includegraphics[width=5.2cm]{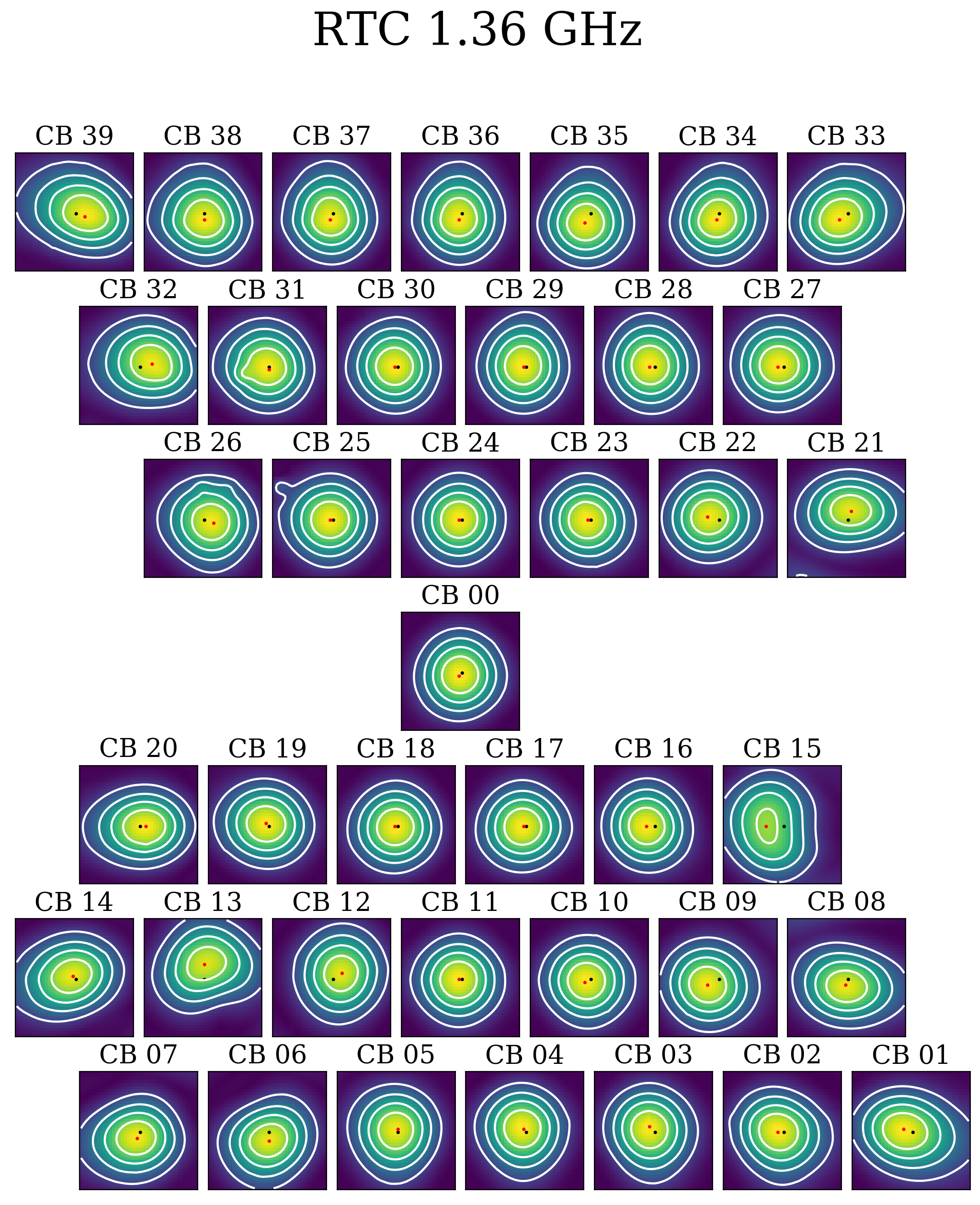}
  \includegraphics[width=5.2cm]{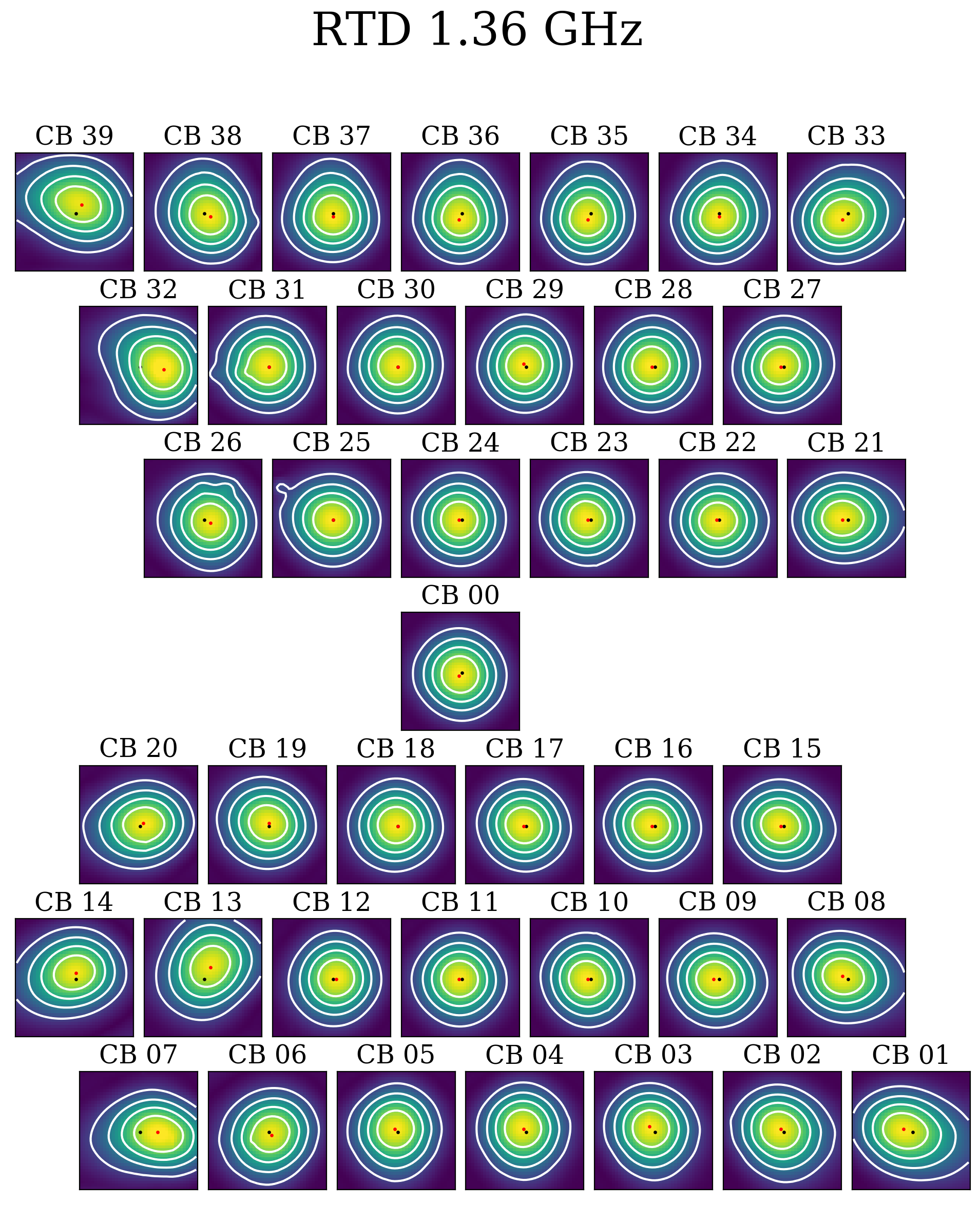}
  \caption{Beam maps for 190821 for individual dishes in I polarisation for the 1.36 GHz frequency bin. The red dot marks the centre of the beam map and the red dot marks the maximum beam response. Contour levels are: 0.2, 0.4, 0.6, 0.8.}
              \label{fig:beams_maps_ant}%
    \end{figure*}

We derive CB maps for individual dishes in the same way as for the whole array. Figure~\ref{fig:beams_maps_ant} shows an example of CB maps at 1.36 GHz derived for individual antennas for the 190821 data set. The plots show that some antennas have distorted CB shapes compared to the same CBs on other antennas, e.g. beam 33 on antenna RT6, beam 18 on antenna RTB, beam 15 on antenna RTC. This is generally due to faulty elements in the PAF on some of the antennas (see also Figure~\ref{fig:elements}). Faulty elements are periodically repaired, which improves the beam shapes. Distorted beams can also be caused by strong RFI during the beam weights measurement. This can especially be a problem for the low frequency part of the observing band. In addition to distortions the maximum sensitivity point of some beams can also be shifted. The degree and direction of this offset varies slightly between the different dishes. This is due to the independent beam weights for the PAFs on the individual dishes.  

The per antenna CB maps can be used with advanced imaging algorithms (e.g. WSClean, \citealt{Offringa2014}) to improve the image quality of the data products. This can be particularly helpful with DDEs caused by broken elements on the PAFs. 

\subsection{Beam size}

\begin{figure}
   \centering
   \includegraphics[width=8.4cm]{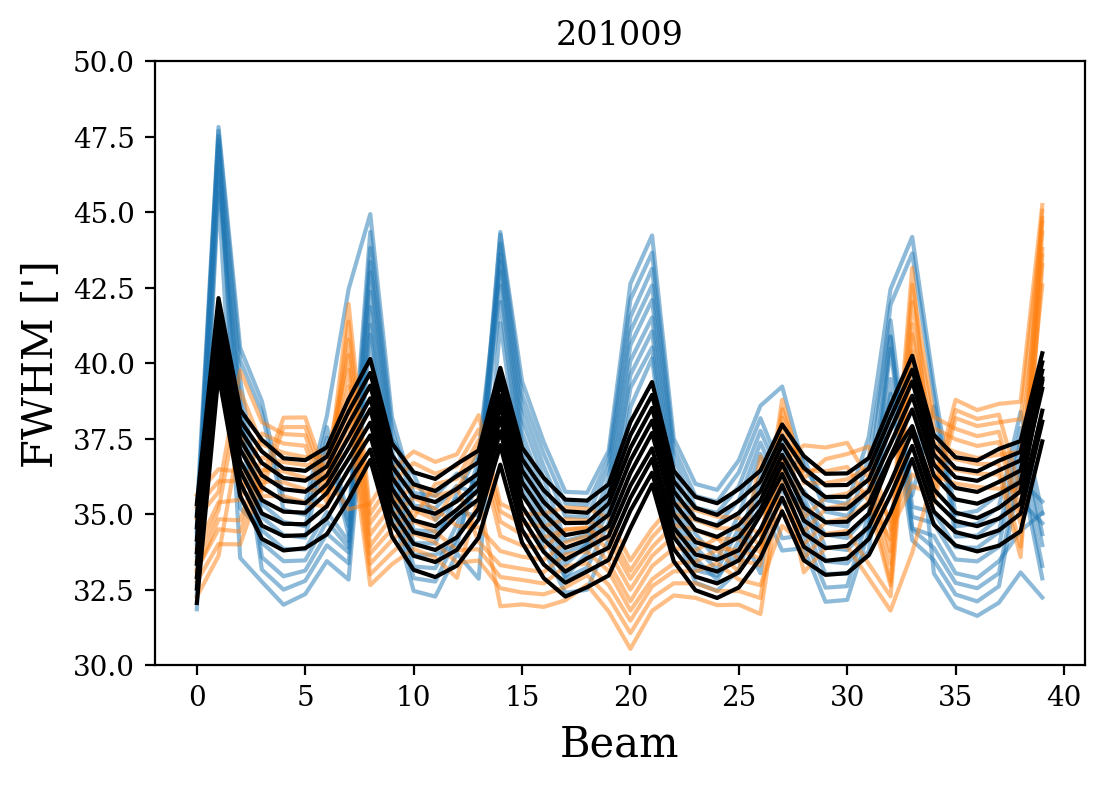}
   \caption{FWHM variation for the 40 CBs. FWHMs are measured by fitting a 2D Gaussian to the beam maps. Blue and orange lines show the FWHM measured along the vertical and horizontal axis and the black line shows the average of the two. There is a line representing each frequency bin for the 201009 data set.}
              \label{fig:FWHM}%
    \end{figure}

To characterise the properties of the CBs we fit a 2D Gaussian to each CB map. Figure~\ref{fig:FWHM} shows the FWHM of the fits as a function of beam number. Blue and orange lines show the FWHM measured along the vertical (Dec) and horizontal (R.A.) axis and the black line shows the average of the two. There is a line representing each frequency bin. The plot shows that CBs in the centre of the field of view (beams 0, 24, 17, 18) are relatively symmetric with similar sizes in the R.A. and Dec direction while the corner CBs (beam 1, 7, 33, 39) are the most asymmetric and diagonally elongated, where the FWHM in different cross sections can differ by $\sim$ 12 arcmin. The elongation of the beams is due to the coma effect and that beams at the edge have fewer contributing elements from the PAF compared to the beams in the centre.    

\subsection{Frequency dependence}

\begin{figure}
   \centering
   \includegraphics[width=8.4cm]{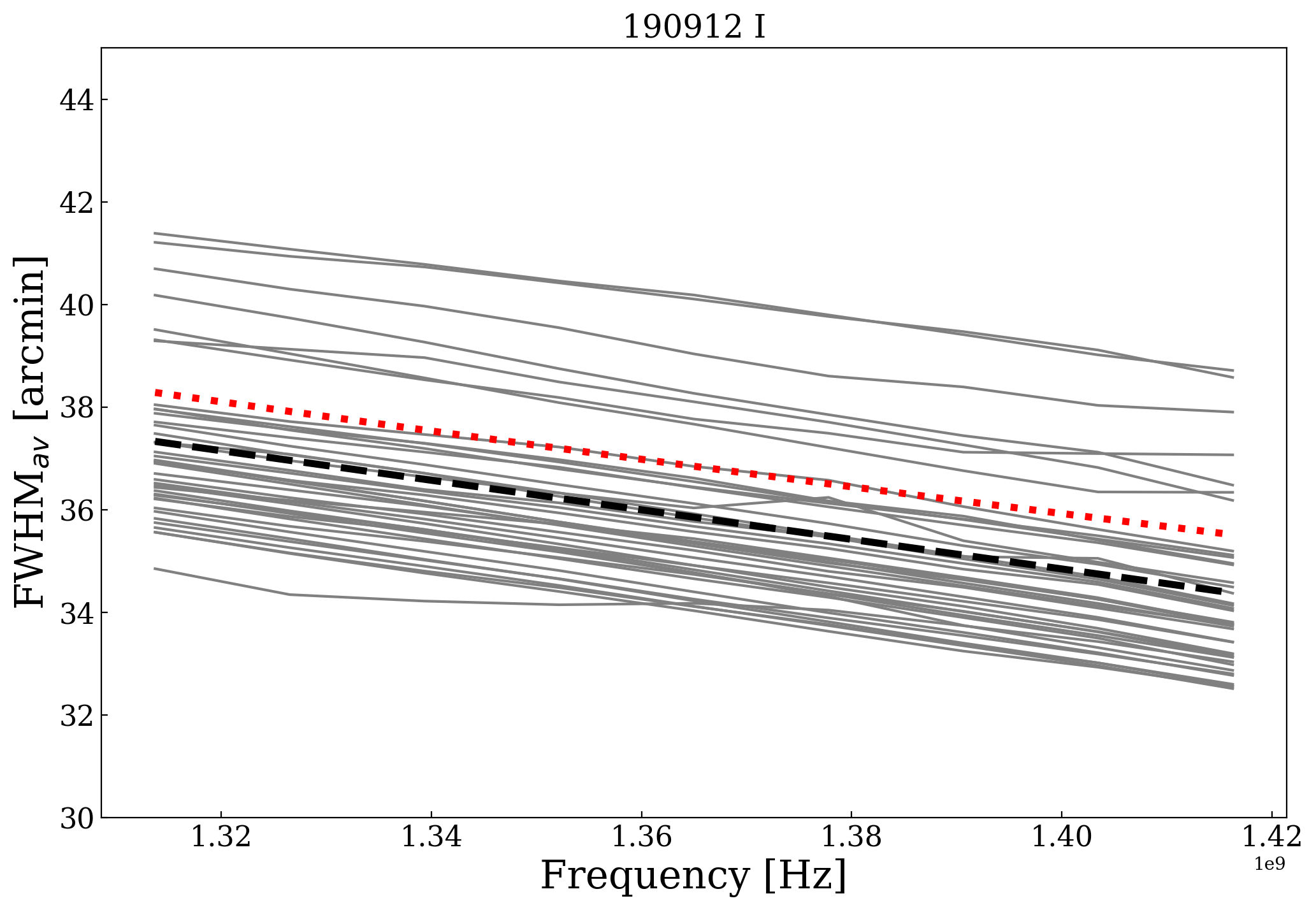}
   \caption{FWHM of CBs as a function of frequency bins. The grey lines show the average FWHM from the 2D Gaussian fit to the I polarisation data of each CB for the data set 190912, while the dashed black line shows the average fitted line and the red dotted line shows the theoretical expectation.} 
              \label{fig:beams_freq}%
    \end{figure}

CB sizes change linearly with frequency. The expected frequency response is $\theta = 1.22$ $\lambda /D$, where $\theta$ is the beam size, $\lambda$ is the wavelength and $D$ is the diameter of the telescope. The measured average frequency dependence for Apertif CBs is:
\begin{equation}
    \theta(\nu) = -3.0 (\pm 0.6) \times 10^{-08} \times \nu + 77 (\pm 8),
\end{equation}
where $\theta$ is the FWHM in arcminutes and $\nu$ is the frequency in Hz. This is based on fitting a 2D Gaussian to each CB map at each frequency, taking the average FWHM from the 2D Gaussian fit and then fitting a first order polynomial to the FWHM vs. frequency. The results were then averaged for all beams in 16 different drift scan measurements. The uncertainty of the parameters is estimated with the standard deviation between the different data sets. The theoretical expected (for a 25m dish) slope and intercept of the relation are: -2.7 $\times 10^{-08}$ and 73.75, the slightly different average measured numbers are due to the more effective illumination of the dish. Figure~\ref{fig:beams_freq} shows the average beam size for each 40 CBs as a function of frequency for a set of drift scans (190912). The dashed black line shows the average fitted line to the data and the red dotted line shows the theoretical expectation. Some of the beams occasionally show non smooth variation with the beam size. In most cases, the cause for this is RFI in certain frequency bins. Beam weights are calculated for each individual subband, which means that the beam shape can be slightly different for each of them, especially if strong RFI is present during the beam weights measurement. Unfortunately, in practice it is not feasible to calculate CB maps per subband, because of the insufficient signal to noise of the data.

\subsection{Time variability}
\label{section:time_variability}

\begin{figure}
   \centering
   \includegraphics[width=8.4cm]{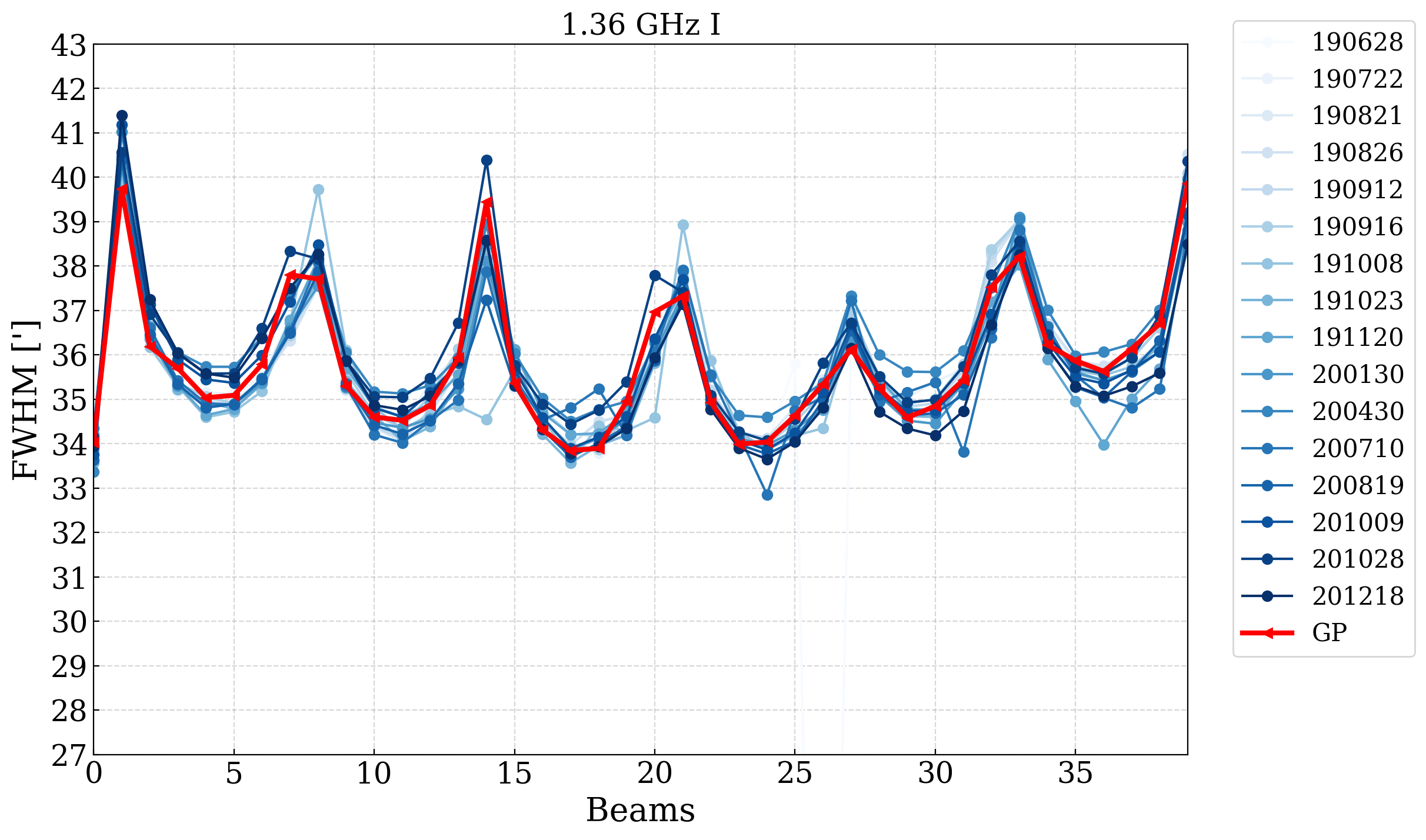}
   \caption{FWHM at 1.35 GHz (chan 5) for all 40 CBs for all data sets up to 201218 (blue lines), including the CB maps derived with the GP method (red line).}
              \label{fig:FWHM_time}%
    \end{figure}

The beam shapes derived from drift scans have average FWHM sizes ranging from 33.5 arcmin to 40.4 arcmin at 1.35 GHz. Over time the average FWHM typically varies by less then 1 arcmin, which is at the few percent level, for CBs in the inner part of the footprint. For CBs at the edge of the footprint, with significantly distorted beam shapes, the average FWHM can vary by up to 2 arcmin ($\sim$ 5\%). Correspondingly the median primary beam correction for the CBs typically varies by a few percent between different drift scan measurements. Figure~\ref{fig:FWHM_time} shows the average FWHM size for 16 different drift scan measurements in 2019 and 2020 at 1.36 GHz. The beam sizes show relatively small variation for most of the beams, $\sim$ 1-2 arc minutes. These can be due to different system conditions, for example broken elements, different RFI conditions, different ambient temperature or small random variations in the electronics of the telescope.  

\subsection{GP CB maps}
\label{section:GP_maps}

The GP CB maps are produced by comparing the flux of continuum point sources in the Apertif survey observations with the measured flux of the same sources in the NVSS survey and then constructing the beam shape with a Gaussian Processes regression method. We use the {\sc scikit-learn gaussian\_process} python library to implement a GP kernel that is a combination of two squared-exponential functions and white noise. One of the squared-exponential functions describes the overall beam shape while the other accounts for small scale deformations. This method requires a certain minimum number of sources for each CB to accurately capture the full CB shape. In practice this means that a minimum of 5 to 7 survey pointings (containing about 500 cross matched continuum sources) are needed to reconstruct the beam shape. There are typically between 6-12 individual pointings observed with the same beam weights. This means, that the GP method can be used to reconstruct the CB shapes for each set of beam weights used for imaging survey observations. For simplicity, the Apertif DR1 is accompanied by a single set of GP CB maps, that are produced from all the high quality observations released in DR1. These maps represent the average CB shapes for the first year of imaging data in DR1. In this work, we present results from this single set of GP CB maps. Considering that we find only a few percent variation in the size of the CBs over this time period (Section~\ref{section:time_variability}), the average GP CB maps are good representations of the Apertif CB shapes. Further details of the calculation of GP beam shapes are discussed in the Apertif DR1 documentation\footnote{The Apertif DR1 documentation: \url{http://hdl.handle.net/21.12136/B014022C-978B-40F6-96C6-1A3B1F4A3DB0} and the GitHub repository of the code: \url{https://github.com/akutkin/abeams}} and in a separate publication \citep{Kutkin_2022}.

\subsection{Comparison of drift scan CB maps to GP CB maps}
\label{section:comparison_to_GP}

\begin{figure*}
  \centering
  \includegraphics[width=8.0cm]{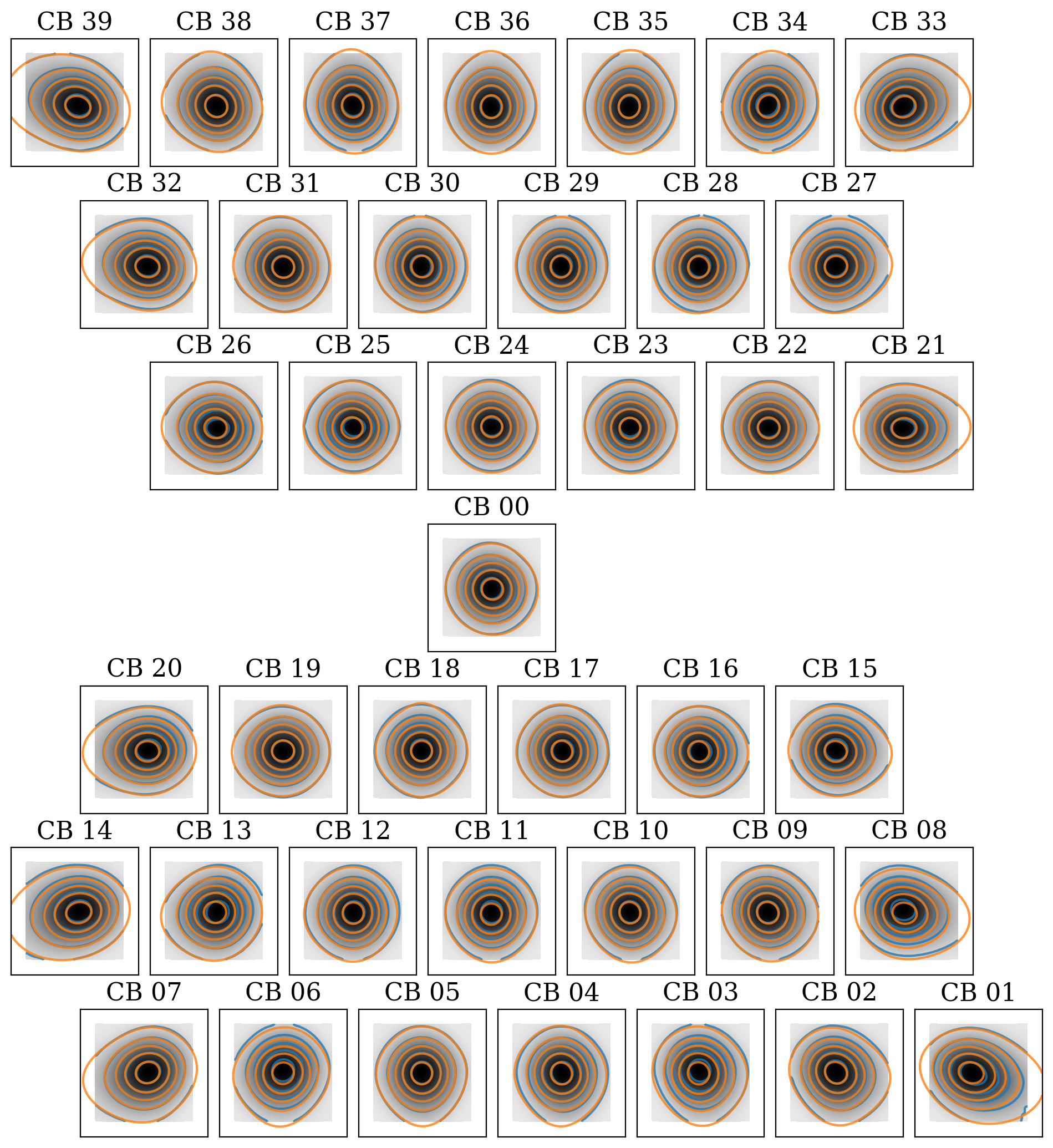}
  \includegraphics[width=8.8cm]{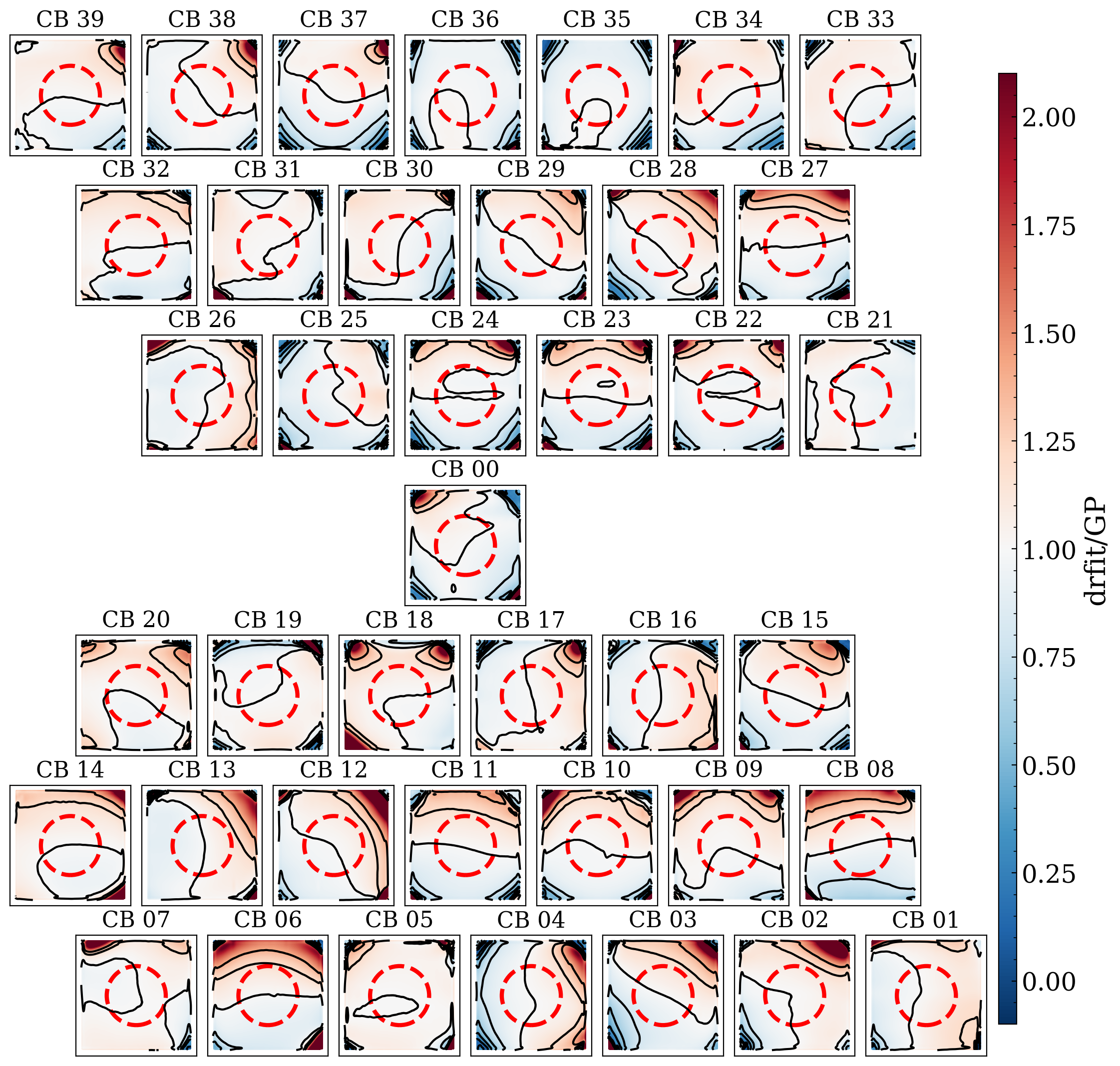}
  \caption{Left:Comparing drift scan CB maps (201028) with GP CB maps. The grey scale map shows the drift scan CB maps. Blue contours show the drift scan data and orange contours show the GP data. Contour levels are the same for both data sets: 0, 0.1, 0.3, 0.5, 0.7, 0.9. Right:Ratio of drift scan CB maps to GP CB maps. Contours are: 0, 0.2, 0.4, 0.6, 0.8, 1, 1.2, 1.4. The red dashed circle marks 36 arcmin from the centre, which indicates the approximate FWHM for the beams.}
              \label{fig:drift_gp_diff}%
    \end{figure*}

\begin{figure}
   \centering
   \includegraphics[width=8.4cm]{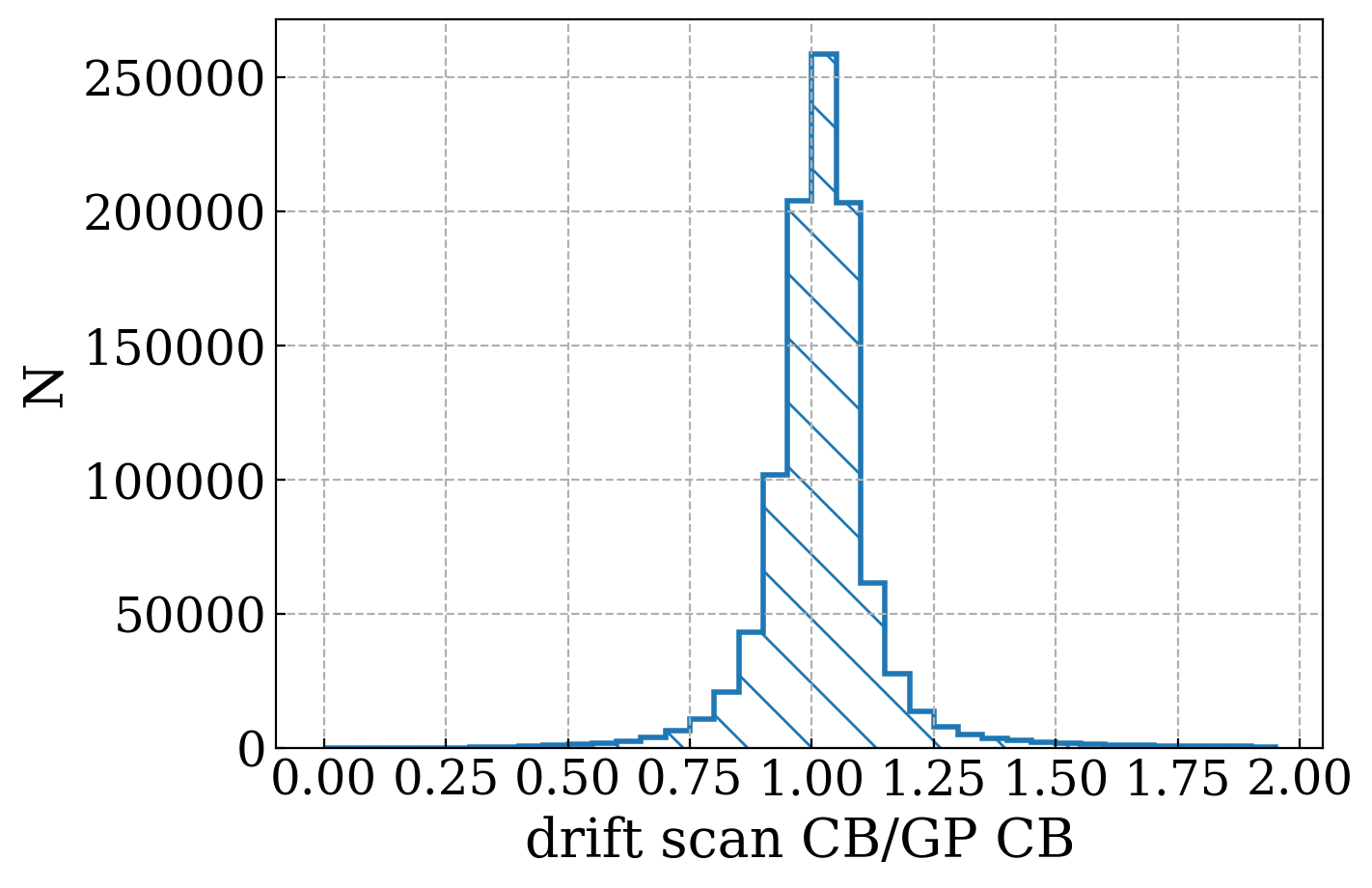}
   \caption{Histogram of the ratio of drift scan CB maps (201028) to GP CB maps.}
              \label{fig:drift_gp_hist}%
    \end{figure}

The comparison of CB maps derived with two independent methods, using drift scans and a GP regression method, gives us the opportunity to better understand the CB shapes of Apertif. Figure~\ref{fig:drift_gp_diff} compares the CB maps derived with the two different methods. The left panel of Figure~\ref{fig:drift_gp_diff} shows a set of drift scan maps in grey scale overlaid with contours of the drift scan data (blue) and the GP map (orange). The two sets of contours match each other very closely, hence for a more detailed comparison the right panel of Figure~\ref{fig:drift_gp_diff} shows the ratio between the drift scan maps (1.36 GHz) and the GP maps. Overall the beam shapes and the average beam sizes agree well between the drift scan CB maps and the GP CB maps. Figure~\ref{fig:drift_gp_hist} shows the distribution of the pixel ratio between the two beam maps (drift scan map (1.36 GHz)/GP), where the drift scan maps are scaled to the same pixel size as the GP maps. The peak of the distribution is 1.03, demonstrating that the two methods produce consistent CB shapes. 

The two methods are highly complementary to each other. The drift scan method has the advantage that it can measure CB shapes for I, XX, and YY polarisation of the Apertif data, for all individual dishes and at several different frequencies. The disadvantages of this method are that the time consuming observing method makes it impractical to measure beam maps for each new set of beam weights and the bright continuum sources that are used for the measurements are not point sources for the individual WSRT dishes. These two things can introduce a bias in the CB maps. The advantages of the GP method are that it only depends on the NVSS catalogue and that CB maps can be derived for each set of beam weight measurements by using the Apertif continuum images observed with the same beam weights within a two week time period. This way the CB maps can reflect the actual status of the system at the time of observing. The disadvantage of the GP method is that it can only be used for the broad band continuum images and can not be derived for different frequencies within the Apertif observing band, for the different polarisation products or for individual dishes. In conclusion, the GP maps can be used for primary beam correcting the multi-frequency continuum images that the Apertif pipeline produces and the drift scan maps can be used in combination with the GP maps to do frequency dependent primary beam correction for spectral line data cubes, or polarisation data cubes; they can be used with advanced imaging algorithms (e.g. \citealt{Offringa2014}) for antenna based deconvolution and primary beam correction; and they can help with identifying problems with Apertif observations, e.g. bad quality CBs on certain antennas at certain times. 

\begin{figure*}
  \centering
  \includegraphics[width=8.4cm]{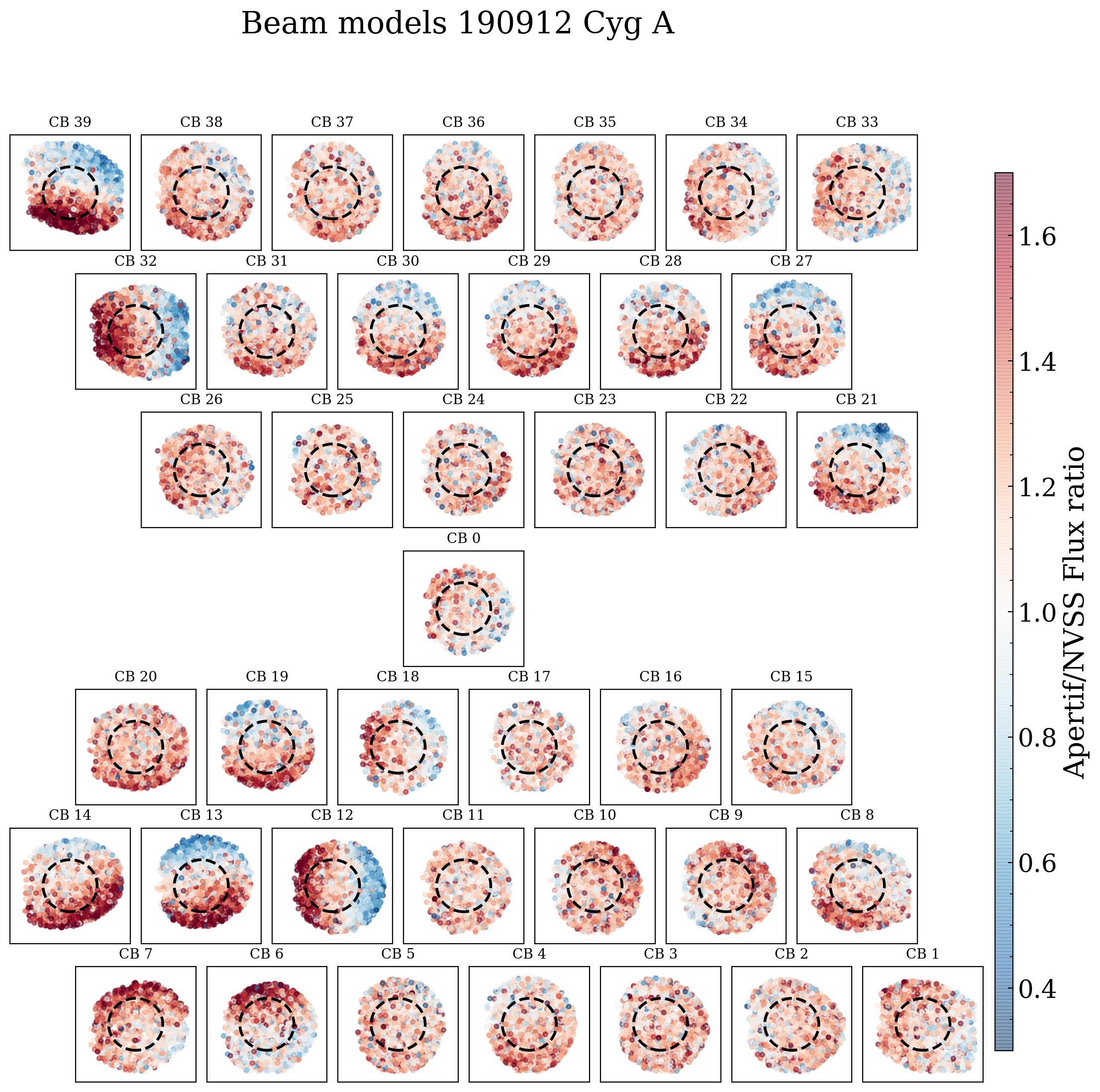}
  \includegraphics[width=8.4cm]{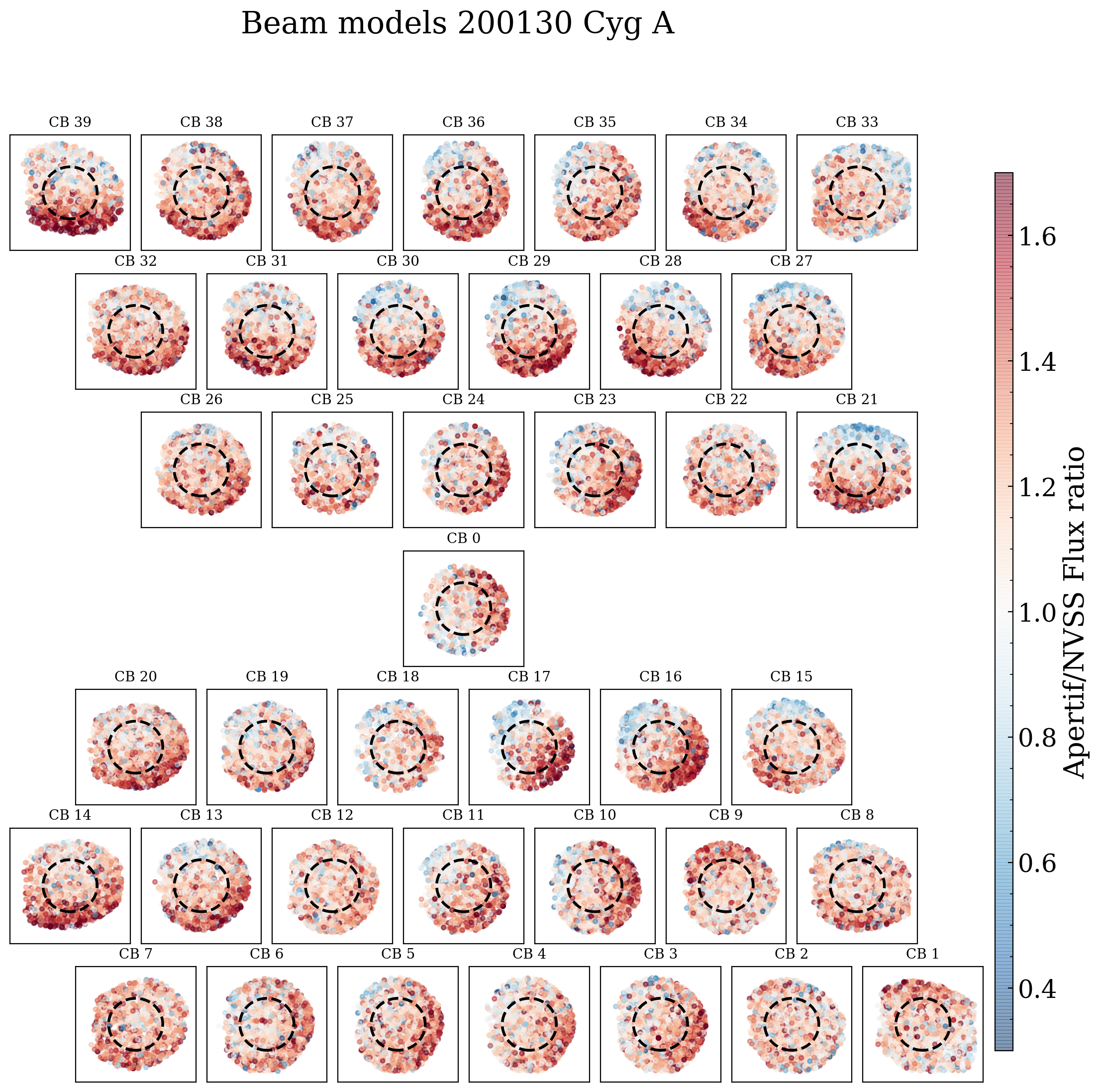}
  \includegraphics[width=8.4cm]{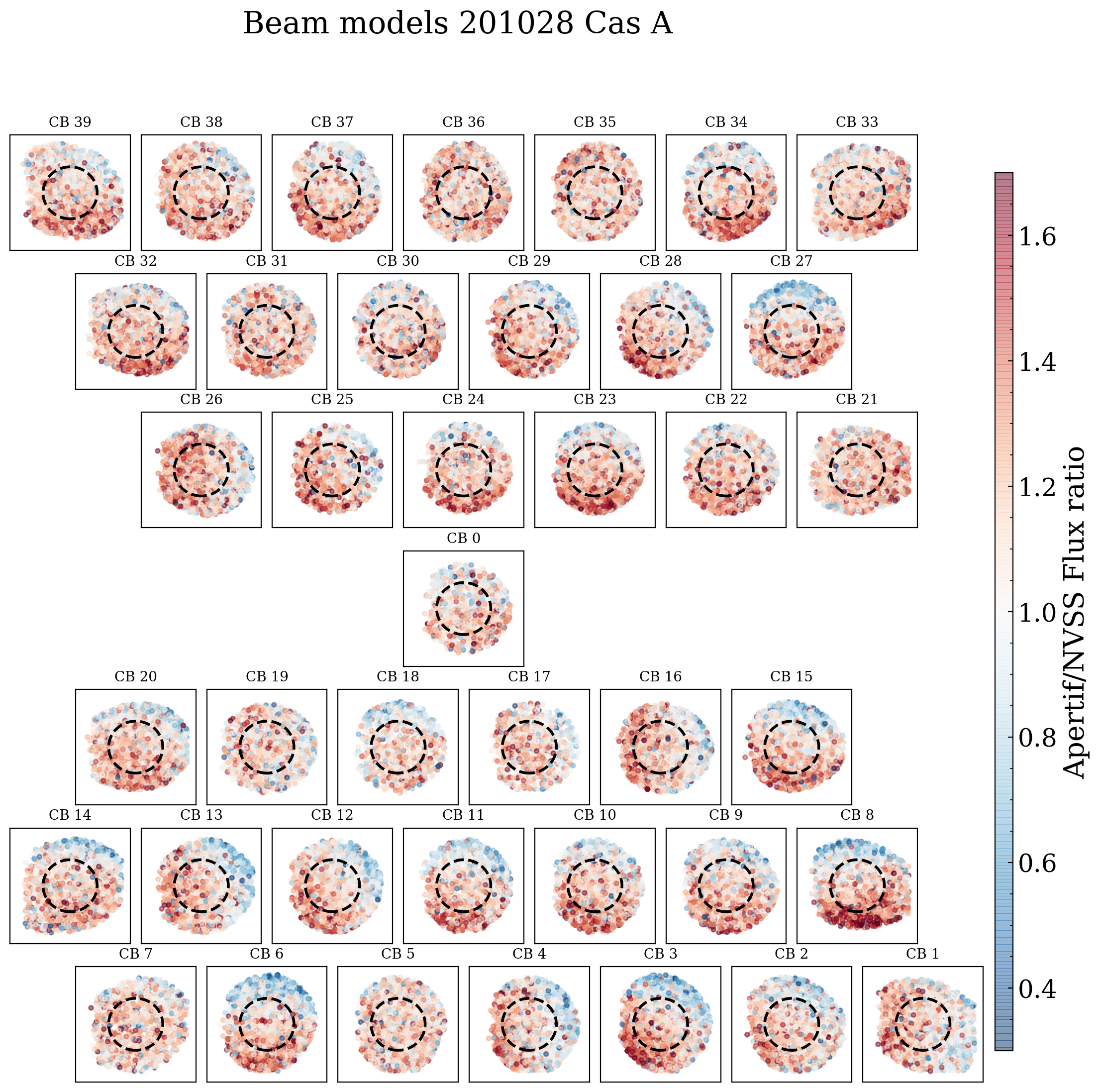}
  \includegraphics[width=8.4cm]{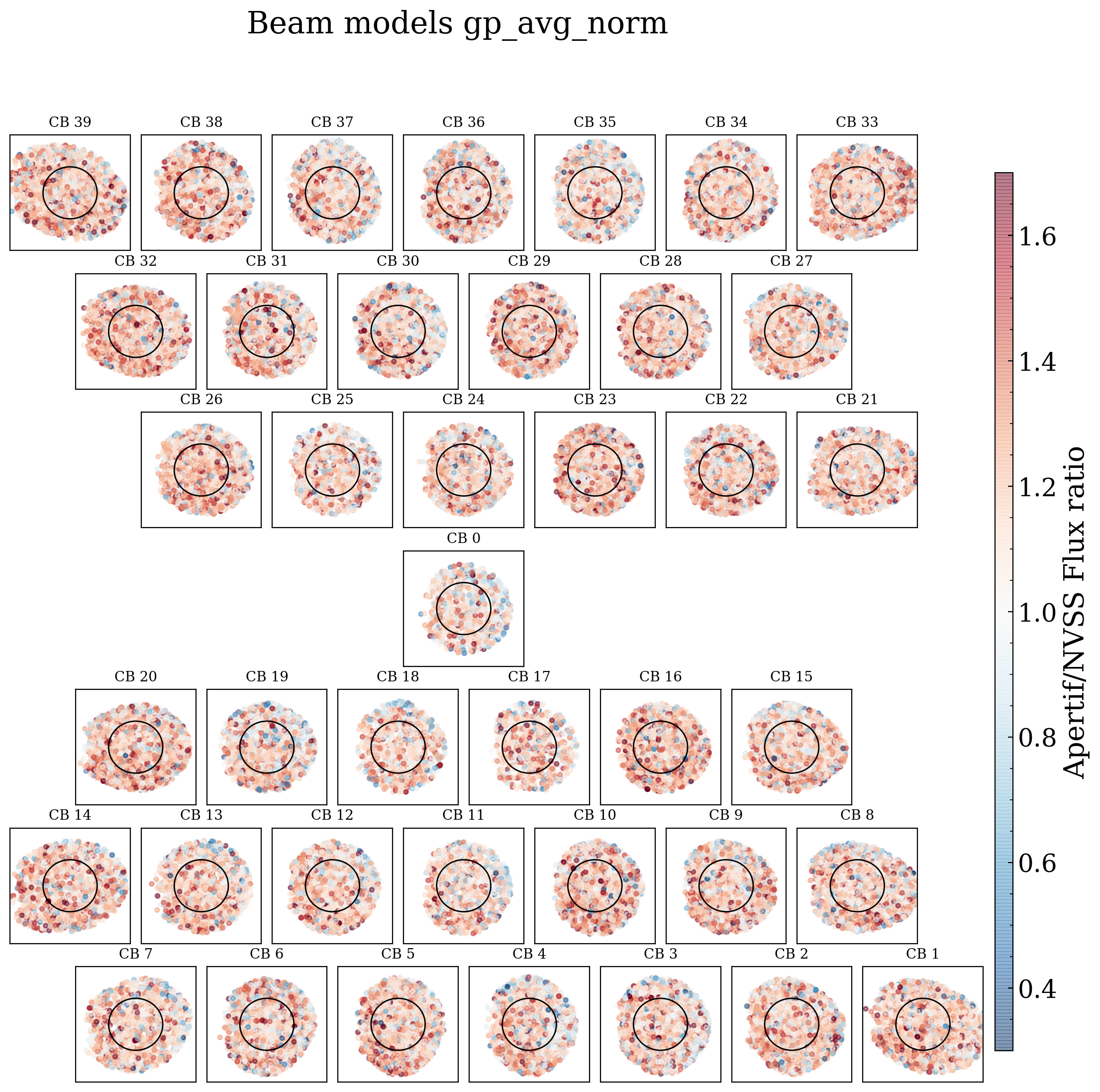}
  \caption{Apertif/NVSS Flux ratio maps with 4 different sets of compound beam maps (190912, 200130, 201028 (Cas A) averaged Gaussian maps). The black dashed circle has a radius of 36 arcmin, which illustrates the average FWHM for the Apertif compound beams. The drift scan based maps clearly show on average higher flux ratios to the south-west side of the beams.}
              \label{fig:flux_ratio_maps}%
    \end{figure*}

To evaluate how representative the different beam maps are for all of the Apertif DR1 data Figure~\ref{fig:flux_ratio_maps} shows maps of Apertif/NVSS flux ratios derived with three different sets of drift scan based CB maps and a set of the average GP CB maps. The plotted flux ratios are derived by first primary beam correcting all continuum images in the Apertif DR1 and then cross matching point sources in the corrected Apertif images with the NVSS catalogue. There is clearly a bias or gradient in the flux ratios for most beams in the drift scan data sets, with an average flux ratio (Apertif/NVSS) of 1.2. The dominant direction of this gradient is different for measurements on Cyg A (higher flux ratio towards the south-west), and on Cas A (higher flux ratio towards the south-east), and the GP beam maps (slightly higher flux ratio towards the south-east). For the GP beam maps this gradient is much milder compared to the drift scan based beam maps. This bias is partially due to that neither CygA nor CasA are point sources. Figure~\ref{fig:sources} shows continuum images of Cas A and Cyg A from the Dwingeloo 820 MHz continuum survey \citep{Berkhuijsen1972}. In addition, both Cyg A and Cas A are close to the Galactic Plane with diffuse large scale continuum emission near both sources. In Figure~\ref{fig:sources} the map of Cyg A shows some extended continuum emission towards the south-west and the map of Cas A has its peak emission towards the south from its nominal coordinates. To try to mitigate this issue we also performed drift scan measurements on 3C 295, which is one of the primary calibrator sources used for Apertif imaging observations, however due to the significantly lower flux ($\sim 22$ Jy) of this source we were not able to produce high quality CB maps from the observations.   

Another cause for the gradient in the Apertif/NVSS Flux ratios in Figure~\ref{fig:flux_ratio_maps} is the quality of the beam weights at the time of the drift scan observations. As discussed in Section~\ref{section:comparison_to_GP} drift scan based CB maps represent the beam shapes for a given beam weights measurement and they depend on the health of the system at a given time. This means that if the quality of certain CBs is not good at the time of the drift scan measurements, these maps would be unsuitable for PB correction for observations with good quality CBs. An example for low quality beam weights are CB 12, 13, 32 and 39 in the top left panel of Figure~\ref{fig:flux_ratio_maps} (beam models 190912). These CB maps show a strong Apertif/NVSS flux ratio bias compared to other CBs in the same drift scan measurements and also compared to the same CBs in other drift scan measurements. This can also account for the smaller gradient in the 200130 data set compared to the 190912 data set, since the quality of the beam weights was substantially better after December 2019 with the improved RFI mitigation strategy. The flux ratio maps also indicate that CB maps from drift scan measurements are generally only suitable for primary beam correction of observations that were taken within a few weeks of the drift scan measurement. On the other hand, science observations with the same low quality CBs will have a larger uncertainty in their measured fluxes when using the GP CB maps for primary beam correction. 

\section{Conclusions}
\label{section:Conclusions}

We use drift scan measurements to map the primary beam response of each of the digitally formed 40 CBs of Apertif. We derive beam maps for all individual beams in I, XX, and YY polarisations in 10 frequency bins. We derive beams maps for the whole Apertif system and also for all 12 individual dishes. We summarise our findings below.

\begin{enumerate}
    \item 
    The beams towards the edge of the Apertif field of view are more asymmetric compared to the centre beams and are diagonally elongated. This effect is the strongest for the corner beams (CB 1, 7, 33 and 39). We wish to emphasise that the use of the classic WSRT primary beam correction is not appropriate for Apertif. 
    \item There is a systematic trend across the Apertif field of view that the CB peak intensity is offset from the nominal beam centres. These offsets correspond to the elongated shape of the CBs and vary slightly between the different dishes. 
    \item We measured Stokes Q leakages on the order of 0.001 at the beam centres and 0.02 at at the 0.5 response level of the CBs. 
    \item We measured the frequency dependence of the Apertif CB sizes. 
    \item We investigated the time variability of the FWHM of the CBs and found that for the inner CBs the standard deviation is typically less than 1 arcmin and for the edge CBs it is approximately 2 arcmin. This corresponds to a 3-5\% variation in the average FWHM of the CBs.
    \item We compared CB maps derived with two independent methods, using drift scans and a Gaussian regression method comparing continuum fluxes measured with Apertif and in NVSS. We find that the overall shape of the CB maps derived with the two different methods agree well.
    \item CBs derived from drift scans are essential to measure the polarisation leakage and the frequency dependence of the CBs. However, GP beams better represent the average shape of the beams since they do not depend on measurements on a single bright continuum source and they can be derived for each individual beam weight setup.
\end{enumerate}

\section{Data availability}

The fits format CB maps generated from the drift scan observations are available from zenodo as two data sets. Frequency setting 1130-1430 MHz DOI: \url{https://doi.org/10.5281/zenodo.6615555},
1220-1520 MHz DOI: \url{https://doi.org/10.5281/zenodo.6544109}. The raw drift scan data are stored on the Apertif Long Term Storage and can be requested for further analysis trough the ASTRON helpdesk (\url{https://support.astron.nl/sdchelpdesk}). The observation taskIDs used for the individual drift scan beam maps are listed within aperPB (\url{https://github.com/apertif/aperPB}, \url{https://doi.org/10.5281/zenodo.6544109}) in the task\_id\_lists directory with the following file names: task\_ids\_\{beamID\}.txt. The GP CB maps are available within the Apertif DR1, trough the ASTRON Virtual Observatory (\url{https://vo.astron.nl/}; detailed documentation on the Apertif DR1 is available here: \url{http://hdl.handle.net/21.12136/B014022C-978B-40F6-96C6-1A3B1F4A3DB0}). 

\begin{acknowledgements}
     We would like to thank the anonymous referee for the useful comments and suggestions, which helped us to improve the paper. This work makes use of data from the Apertif system installed at the Westerbork Synthesis Radio Telescope owned by ASTRON. ASTRON, the Netherlands Institute for Radio Astronomy, is an institute of the Dutch Research Council (“De Nederlandse Organisatie voor Wetenschappelijk Onderzoek, NWO). KMH acknowledges financial support from the State Agency for Research of the Spanish Ministry of Science, Innovation and Universities through the ``Center of Excellence Severo Ochoa'' awarded to the Instituto de Astrof\'isica de Andaluc\'ia (SEV-2017-0709) from the coordination of the participation in SKA-SPAIN, funded by the Ministry of Science and innovation (MICIN). EAKA is supported by the WISE research programme, which is financed by the Netherlands Organization for Scientific Research (NWO). JMvdH and KMH acknowledges funding from the European Research Council under the European Union’s Seventh Framework Programme (FP/2007-2013)/ERC Grant Agreement No. 291531 (‘HIStoryNU’). BA acknowledges funding from the German Science Foundation DFG, within the Collaborative Research Center SFB1491 ”Cosmic Interacting Matters - From Source to Signal”. YM, LCO, and RS acknowledge funding from the European Research Council under the European Union's Seventh Framework Programme (FP/2007-2013)/ERC Grant Agreement No. 617199. JvL acknowledges funding from Vici research programme `ARGO' with project number 639.043.815, financed by the Dutch Research Council (NWO). DV acknowledges support from the Netherlands eScience Center (NLeSC) under grant ASDI.15.406. This research has made use of {\sc NumPy} \citep{vanderWalt2011}, {\sc scipy} \citep{2020SciPy}, {\sc Astropy}, a community-developed core {\sc Python} package for Astronomy (\citealt{astropy:2013, astropy:2018}; http://www.astropy.org), and {scikit-learn} \citep{scikit-learn}. 
\end{acknowledgements}

%
\bibliographystyle{aa} 
\bibliography{apertif_beams} 
%
\begin{appendix} 
\section{List of compound beam maps}

\begin{table*}
\caption{List of high quality CB maps. The first column shows the CB map ID, the second column denotes the continuum source used for the drift scans, the third column denotes the observing frequency range and the fourth column contains notes about the observations, for example missing antennas, significant issues with RFI or beam weights.}            
\label{table:observations}      
\centering                          
\begin{tabular}{c c c c}        
\hline\hline                 
CB map ID & source & frequency range  & notes \\    
 &  & [MHz] &  \\    
\hline                        
  190628 & Cyg A & 1130-1430 & issue in channel 2 (1.31 GHz) with CBs 15-16,\\
  & & & issue in channels 5-6 (1.36 - 1.37 Ghz) and 8-9 (1.4 - 1.41 GHz)\\ 
  190722 & Cyg A & 1130-1430 & RFI in chan 6 (1.37 GHz) for CB 1 \\
  190821 & Cyg A & 1130-1430 & \\
  190826 & Cyg A & 1130-1430 & \\
  190912 & Cyg A & 1130-1430 & \\
  190916 & Cyg A & 1130-1430 & \\
  191008 & Cyg A & 1130-1430 & CBs 14, 19, 20 and 26 are distorted \\
  191023 & Cyg A & 1130-1430 & RFI in channel 6 ( 1.37 GHz) for CBs 0, 3-4, 9-10, 15-17, 22 \\
  191120 & Cyg A & 1130-1430 & RFI in channels 8-9 (1.4 - 1.41 GHz)\\
  200130 & Cyg A & 1130-1430 & no RT5, RFI in channel 6 (1.37 GHz) for CBs 1, 6-8, 14, 21, 33 \\
  200430 & Cyg A & 1130-1430 & no RTC and RTD \\
  200710 & Cyg A & 1130-1430 & no RTB, CBs 0, 20-39 are affected by RFI\\
  200819 & Cyg A & 1130-1430 & issue with CBs 37-39, RFI in channel 6 (1.37 GHz)\\
  201009 & Cyg A & 1130-1430 & no RTB \\
  201028 & Cas A & 1130-1430 & \\
  201218 & Cyg A & 1130-1430 & \\
  210205 & Cyg A & 1220-1520 & RFI in channels 13-15 (1.46-1.48 GHz)\\
  210402 & Cas A & 1220-1520 & RFI in channels 13-15 (1.46-1.48 GHz), no RTC and RTD\\
  211103 & Cas A & 1220-1520 & RFI in channels 13-15 (1.46-1.48 GHz)\\
  211118 & Cyg A & 1220-1520 & RFI in channels 13-15 (1.46-1.48 GHz)\\
\hline                                   
\end{tabular}
\end{table*}

\end{appendix}

\end{document}